\documentclass[pre, reprint,twocolumn,showpacs,preprintnumbers,amsmath,amssymb,a4paper]{revtex4-1}
\usepackage{graphicx}
\usepackage{epsfig}
\usepackage{natbib}
\usepackage{amsfonts}
\usepackage{wasysym}
\providecommand{\abs}[1]{\lvert#1\rvert}

\begin{document}

\title{Formation of localized structures in bistable systems through nonlocal
spatial coupling II: The nonlocal Ginzburg Landau Equation}

\author{Lendert Gelens$^{1,2}$, Manuel A.~Mat\'{\i}as$^2$, Dami\`a Gomila$^{2}$, Tom Dorissen$^1$,
and Pere Colet$^2$}
\affiliation{
$^1$Department of Applied Physics and Photonics, Vrije Universiteit Brussel,
Pleinlaan 2, 1050 Brussels, Belgium;\\
$^2$IFISC, Instituto de F\'{\i}sica Interdisciplinar y Sistemas 
Complejos (CSIC-UIB), Campus Universitat Illes Balears, E-07122 Palma de
Mallorca,  Spain}

\date{\today}


\begin{abstract}
We study the influence of a linear nonlocal spatial coupling on the interaction of fronts connecting two equivalent stable states in the prototypical $1$-D real Ginzburg-Landau equation. While for local coupling the fronts are always monotonic and therefore the dynamical behavior leads to coarsening and the annihilation of pairs of fronts, nonlocal terms can induce spatial oscillations in the front, allowing for the creation of localized structures, emerging from pinning between two fronts. We show this for three different nonlocal influence kernels. The first two, mod-exponential and Gaussian, are positive-definite and decay exponentially or faster, while the third one, a Mexican-hat kernel, is not positive definite.
\end{abstract}

\maketitle

\section{Introduction}
\label{Sect::Introduction}

We have shown recently \cite{GelensPRL2010} that a nonlocal interaction term can induce oscillatory tails in otherwise monotonic fronts connecting two equivalent homogeneous steady states (HSSs) in the Ginzburg-Landau equation (GLE) for a 1-dimensional (1-D) real field. As a consequence the interaction between a pair of fronts has an oscillatory dependence with the distance between the fronts with an exponentially decaying envelope. The oscillatory dependence allows for a pair of fronts to be pinned at specific distances determined by the tail profile. In particular localized structures (LSs) can arise as a consequence of the pinning.

In \cite{PartI}, which we will refer to as Part I here, we have presented a suitable framework to understand the effect of linear nonlocal spatial coupling on the shape of a class of fronts connecting two equivalent HSSs allowing to determine the parameter regions where fronts have an oscillatory profile. Here we apply this general framework to rationalize and extend the results advanced in \cite{GelensPRL2010} and elucidate the region in parameter space where LSs can exist for different forms of nonlocal interaction. 

In particular in Part I we considered 1-D extended systems described by a real field with a nonlocal interaction term $s F(x,\sigma)$, in which $s$ is a parameter that controls the overal strength and sign of the coupling while $F(x,\sigma)$ can be written as the convolution of a spatially nonlocal kernel $K_{\sigma}(x)$, with the field $A(x)$
\begin{equation}
F(x,\sigma) =  \int_{-\infty}^{\infty} \! K_{\sigma}(x-x')\,A(x')dx'\ ,
\label{Eq::nonlocal_term}
\end{equation}
where the parameter $\sigma$ controls the spatial extension (width) of the coupling.

\begin{figure}
\begin{center}
\includegraphics[width=8.6cm]{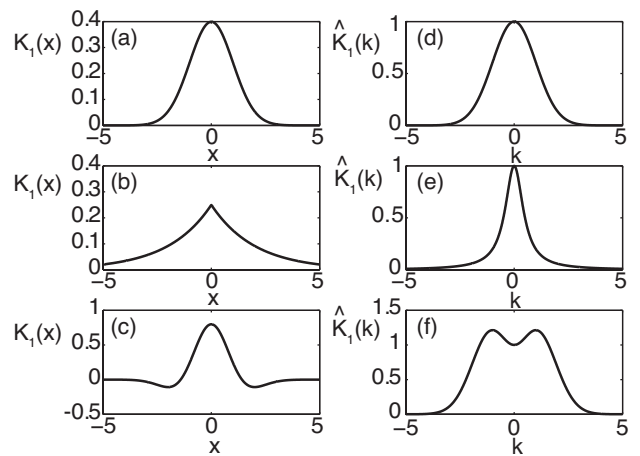}
\caption{\label{Fig::kernel_shapes} Representations of the three nonlocal interaction kernels used in this work ($\sigma=1)$:
(a) Gaussian; (b) mod-exponential (Laplacian); (c) Mexican-hat ($b=1/2$). The right panels (d)-(f) show the Fourier transforms 
of the kernels in panels (a)-(c), respectively.}
\end{center}
\end{figure}

Here we will consider three different interaction kernels that 
illustrate the generality of the spatially nonlocal effects considered. Two of the kernels are positive definite, Gaussian and mod-exponential, while the third is a non-positive definite kernel, a Mexican-hat kernel. Table \ref{tab:kernels} gives the expression in real and Fourier space of the three kernels used in the present work, while Fig.~\ref{Fig::kernel_shapes} displays its shape. Moreover, these three kernels are relevant in different applications.

\begin{table*}
	\centering
		\begin{tabular}{lllll}
		  & & \textbf{Gaussian} & \textbf{mod-exponential} & \textbf{Mexican hat} \\
		\hline
		$K_{\sigma}(x)$: && $\frac{1}{ \sqrt{2 \pi} \sigma} e^{-x^2/ 2 
\sigma^2}$ &  $\frac{1}{ 4 \sigma} e^{- \abs{x}/ 2 \sigma}$ & $\frac{\sqrt{2}}{\sqrt{\pi} \sigma} 
\left(1- b \frac{x^2}{\sigma^2}\right) e^{-x^2/ 2 \sigma^2}$ \\	
		 $\hat{K}_{\sigma}(k)$:&&   $e^{-k^2 \sigma^2 / 2 }$ & $\frac{1}{1+4 
\sigma^2 k^2}$ & $ 2 (1+ b (-1 + \sigma^2 k^2) ) e^{-\sigma^2 k^2/ 2 }$  \\
		  \\
		\end{tabular}
		\caption{Definition of the three different kernels $K_{\sigma}(x, \sigma)$ used in this work together with their Fourier 
transform $\hat{K}_{\sigma}(k, \sigma)$: Gaussian, the mod-exponential and Mexican-hat kernels.}
	\label{tab:kernels}
\end{table*}

Spatially nonlocal interactions in which the spatial interaction kernel is positive definite are either attractive (activatory) or repulsive (inhibitory) depending on the sign of $s$. The most usual kernels of this type
are the Gaussian and the exponential, that have been studied in several contexts like competition effects in Ecology \cite{Emilio_2009,*Emilio_2010,Clerc_PRE_2005,Clerc_PRE_2010}, Nonlinear Optics \cite{Krolikowski_PRE_2001, *Wyller_PRE_2002, *Ouyang_PRE_2006}, reaction-diffusion systems \cite{Bordyugov_PRE_2006} and Neuroscience \cite{Ermentrout_RPP, Hellwig,Coombes05}. Physically, in reaction-diffusion systems an spatially nonlocal interaction with an exponential kernel arises whenever an adiabatic elimination of a fast diffusing substance is performed \cite{Shima_PRE_2004}. More general exponential decaying kernels with a variable exponent, including the exponential and Gaussian as particular cases, have been also considered \cite{Emilio_2009,*Emilio_2010}. In some instances it is possible to reconstruct an interaction kernel from experimental data, as it is done in Ref. \cite{Minovich} for a thermal nonlinear optical medium.

Non monotonic kernels were introduced in the context of neuroscience. 
Neurons are intrinsically discrete units, but one can make use of continuous neural field models
that are coarse grained description of the spatiotemporal evolution at the tissue level. 
These models, like the two-layer (one activatory and the other inhibitory) network
Wilson-Cowan model \cite{WilsonCowan1972} typically include nonlocal effects.
A Mexican-hat kernel was introduced by Amari \cite{Amari1977} to describe in terms of an effective field model a mixed population of activatory and inhibitory neurons.
Since then, Mexican-hat kernels displaying local activation and lateral inhibition characteristics, have been used extensively in Neuroscience \cite{Ermentrout_RPP,Coombes05}
and also in reaction-diffusion \cite{Hildebrand2001}. The reverse situation, local inhibition
and lateral excitation has been also considered \cite{Ermentrout_RPP}. Here
we will study a kernel resulting from the combination
of two Gaussians as shown in Table I and Fig. 1(c) which also allows
to recover the results for a Gaussian kernel in the limit
$b\rightarrow0$. 

The manuscript is organized as follows. In Section \ref{Sect::GLE} we briefly describe the GLE equation with nonlocal interaction. In Section \ref{Sect::moments} we discuss the moment expansion of the kernel. Sections
\ref{Sect::gauss}, \ref{Sect::expon} and \ref{Sect::Mexhat} are devoted to the Gaussian, mod-exponential and Mexican-hat kernels respectively. Finally concluding remarks are given in \ref{Sect::conclu}.

\section{The Ginzburg-Landau equation with nonlocal interaction}
\label{Sect::GLE}

The prototypical cubic Ginzburg-Landau equation (GLE) for a real field $A$ in 1 spatial dimension can be written as \cite{Cross_RevModPh_1993},
\begin{equation}
\partial_t A  = \mu  A - A^3 + \partial_{xx} A \,  .
\label{eq:gle}
\end{equation}
The parameter $\mu$ is the gain coefficient. The coefficients of the diffusion and cubic terms are set to one by suitably rescaling the spatial and temporal scales without loss of generality. The GLE is symmetric under the parity transformation $x\leftrightarrow -x$. 

For $\mu<0$ the origin, $A_s=0$, is the only steady state (stable) in the system. At $\mu=0$ the system exhibits a pitchfork bifurcation, and two stable, symmetry related, HSSs appear at $A_s=\pm\sqrt{\mu}$. For $\mu>0$ the system is bistable and exhibits front solutions (kinks and anti-kinks) that connect the two HSSs. The fronts always decay to the HSS in a monotonic way. 

We now consider an additional nonlocal term $F(x,\sigma)$ defined as in Eq.~(\ref{Eq::nonlocal_term}). We assume that the kernel is real and preserves the symmetry under the parity transformation $x\leftrightarrow -x$, namely $K_{\sigma}(x)=K_{\sigma}(-x)$. The extension of the GLE with nonlocal coupling can then be written as
\begin{equation}
	\partial_t A  = (\mu - sM_0) A - A^3 +  
	\partial_{xx} A  + 
	s  F(x,\sigma)  ,
\label{Eq::RGL}
\end{equation}
where $s$ determines the strength of the nonlocal term. The term $-s M_0 A$ where $M_0 = \int_{-\infty}^{\infty} K_{\sigma}(x) dx$ compensates for the local contribution of $F(x,\sigma)$. Through this compensation the nonlocal system (\ref{Eq::RGL}) has the same HSSs as the GLE with local coupling (\ref{eq:gle}). 

To analyze the linear stability of a HSS we consider perturbations of the from $A = A_s + \epsilon \exp{(\Gamma t + i k x)}$. Linearizing for small $\epsilon$ one obtains for Eq. (\ref{Eq::RGL}) the dispersion relation:
\begin{equation}
\Gamma (k) = \mu' -k^2 + s (\hat{K}_{\sigma}(k)-M_0) \, ,
\label{Eq::HSS_stab_k}
\end{equation}
where 
\begin{equation}
\hat{K}_{\sigma}(k) =  \int_{-\infty}^{\infty} \! K_{\sigma}(x)\,e^{-ikx} dx , 
\label{eq:Kernel_Fourier}
\end{equation}
is the Fourier transform of the kernel. Owing to the kernel symmetries $\hat{K}_{\sigma}(k)=\hat{K}_{\sigma}(-k)$ and the dispersion relation $\Gamma(k)$ depends on $k$ only through $k^2=u$ and we can write
\begin{equation}
\tilde \Gamma (u) = \mu' - u + s (\tilde{\hat{K}}_{\sigma}(u)-M_0) \, ,
\label{Eq::HSS_stab_u}
\end{equation}
where $\mu^{\prime}=\mu-3 A_s^2$. For $\mu\le 0$, $A_s=0$ and $\mu^{\prime}=\mu$ while for $\mu> 0$, $A_s=\pm\sqrt{\mu}$ and $\mu^{\prime}=-2\mu<0$. As a consequence all stable steady states of the GLE are associated to a negative value for $\mu^{\prime}$.

A given HSS becomes unstable if the maximum of $\Gamma(k)$ becomes positive at some $k_c$. If $k_c=0$ the instability is associated to a homogeneous perturbation while if $k_c \neq 0$ the system undergoes a modulational instability (MI).
For the local GLE, $s=0$, the dispersion relation has a parabolic shape with a single maximum at $k=0$ where $\Gamma(0)=\mu'$.  Changing the parameter $\mu$ the parabola moves rigidly in the vertical direction, so the maximum of the dispersion relation is always located at zero and therefore none of the HSSs can undergo a MI. As for homogeneous instabilities, the zero HSS becomes unstable at $\mu=0$, where the two steady states at $A_s=\pm\sqrt{\mu}$ are born (pitchfork bifurcation). The two non-zero
symmetric HSSs are always stable in the parameter region where they exist. As we will see later, nonlocality induces a MI if, for some $k_c$, the last term in eq.~(\ref{Eq::HSS_stab_k}) overcomes the stabilizing $\mu^{\prime}-k^2$ terms, making $\Gamma(k_c)=0$ for a finite $k_c$.

We focus now on stationary spatial structures. Setting the time derivative to zero in (\ref{Eq::RGL}) one has, 
\begin{equation}
 \partial_{xx} A = (- \mu + s M_0) A + A^3 - s  F(x,\sigma)  \, .
\end{equation}
Defining the intermediate variable $V$, one obtains the following 2-dimensional {\it spatial} dynamical system
\begin{align}
A' &=V \nonumber\\
V' &=(-\mu +sM_0)A+A^3 -s F(x,\sigma) \ ,
\label{eq:spatdynsys} 
\end{align}
where the prime stands for derivatives with respect to the spatial variable $x$. The fixed points of (\ref{eq:spatdynsys}) corresponds to solutions for $A$ which do not depend on $x$, thus to HSSs. Close to a HSS the shape of the fronts starting or ending at it can be obtained by considering a perturbation of the form $A(x) = A_s + \epsilon \exp(\lambda x)$ (where, in general, $\lambda$ is complex) and linearizing for small $\epsilon$. The spatial eigenvalues fulfill 
\begin{equation}
 \Gamma_s(\lambda)=0
\end{equation}
where $\Gamma_s(\lambda)$ is the dispersion relation (\ref{Eq::HSS_stab_k}) replacing $k$ by a complex $-i\lambda$, namely
\begin{equation}
 \Gamma_s(\lambda) = \Gamma (-i\lambda) = \mu' + \lambda^2 + s (\hat{K}_{\sigma}(-i\lambda)-M_0) \, .
\end{equation}
$\Gamma_s(\lambda)$ depends on $\lambda$ only through $\lambda^2=-u$, thus spatial eigenvalues can also be obtained from $\tilde \Gamma(u_0)=0$. If $u_0$ is real then there is a doublet of spatial eigenvalues $\lambda_0=\pm\sqrt{-u_0}$ with $\lambda_0$ real for $u_0<0$ or purely imaginary for $u_0>0$. If $u_0$ is complex then $u_0^*$ is also a zero and therefore complex spatial eigenvalues come one in quartets $\lambda_0=\pm q_0 \pm i k_0$. 

As discussed in Part I \cite{PartI}, if the eigenvalues are well separated the leading eigenvalues, i.e. those with the smallest real part, determine the asymptotic approach to the HSS. If the leading eigenvalues are a real doublet, fronts approach monotonically to the HSS. If the leading eigenvalues are an imaginary doublet, the HSS is modulationally unstable. The most interesting case is when the leading eigenvalues are a complex quartet, since fronts starting or ending at the HSS have oscillatory tails and thus LSs may arise as a consequence of the tail interaction. 

Varying parameters two doublets can collide and lead to a complex quartet and viceversa. As explained in Part I the collision is signaled by a real double zero (RDZ) of $\tilde \Gamma (u)$, namely $\tilde \Gamma(u_c)=\tilde \Gamma'(u_c)=0$ for $u_c \in R$. If $u_c>0$ two imaginary doublets become a complex quartet signaling a Hamiltonian-Hopf (HH) bifurcation. If the HH occurs at the maximum of $\Gamma(k)$ it corresponds to a MI. If $u_c<0$ two real doublets become a complex quartet which corresponds to the so-called Belyakov-Devaney (BD) transition \cite{Devaney76}. Moving on top of the RDZ manifold $u_c$ changes value and eventually can change sign, so that a MI becomes a BD and viceversa. This happens when $\tilde \Gamma (0)=\tilde \Gamma' (0)=0$ and corresponds to a quadruple zero (QZ) of the dispersion relation when written as function of $k$, $\Gamma(k)$ \cite{PartI}.

As shown in Part I, besides the QZ, there are two other codim-2 points that play a relevant role in organizing the phase space dynamics, namely the cusp point where two BD or two HH manifolds start (or end) and the 3DZ$(i\omega)$ in which the HSS becomes simultaneously unstable to homogeneous and finite wavelength perturbations.

\section{Moment expansion} \label{Sect::moments}

A qualitative understanding of the effects of a nonlocal coupling by means of a moment expansion is possible for
kernels that in Fourier space have no singularities at finite distances, as is the
case of the Gaussian kernel to be considered in Sec.~\ref{Sect::gauss}. Proceeding as indicated in Part I, the nonlocal interaction can be written as a series of spatial derivatives of $A$ (see also \cite{Murray_Springer_2002})
\begin{equation}
F(x,\sigma)=  \sum_{j=0}^{\infty} \frac{M_{2j}}{(2j)!}  \frac{\partial^{2j} A}{\partial x^{2j}} \, ,
\label{eq::taylorexp}
\end{equation}
where $M_j = \int_{-\infty}^{\infty} x^j K_{\sigma}(x) dx$. To describe the BD and MI transitions which involve 4 spatial eigenvalues we need to keep the expansion terms at least up to fourth order derivatives.

For the GLE (\ref{Eq::RGL}) truncating the expansion of the kernel at fourth order one has
\begin{equation}
	\partial_t A = \mu A - A^3 +  
	\left(1+ \frac{1}{2!}s M_2\right)\nabla^2 A  + 
	\frac{1}{4!} s M_4 \nabla^4 A \, .
\label{Eq::momexpan} 
\end{equation}
Eq.~(\ref{Eq::momexpan}) is related to the widely studied Swift-Hohenberg equation \cite{SH1977,*MalomedNT90,*Bestehorn90,BurkeK07}. 
An analogous truncation for a spatially nonlocal interaction was considered in
Ref. \cite{Gelens_PRA_2007,*Gelens_PRA_2008}. Notice Eq.~(\ref{Eq::momexpan}) only makes sense if $sM_4<0$ since otherwise large wavenumber perturbations will always be amplified leading to divergences. Table~\ref{tab:moments} gives the values of the moments for the non-singular kernels considered in this article. 

\begin{table}
	\centering
		\begin{tabular}{llll}
		\textbf{Moment~~} & \textbf{Gaussian~~}  & \textbf{Mexican hat} \\
		\hline
		$M_0$ & 1 & $2 (1 - b)$\\
		$M_2$ & $\sigma^2$ &  $2 (1 -  3b) \sigma^2$ \\
		$M_4$ & $3 \sigma^4$ &  $6 (1 -  5 b) \sigma^4$\\
		$M_6$ & $15 \sigma^6$ &   $30 (1 -  7 b ) \sigma^6$\\
		$M_8$ & $105 \sigma^8$ &   $210 (1 -  9 b ) \sigma^8$\\				
		\end{tabular}
	\caption{First moments $M_i$ for the Gaussian and the Mexican-hat kernels}
	\label{tab:moments}
\end{table}

The dispersion relation for the $4$th order truncated moment expansion is,
\begin{equation}
\tilde{\Gamma}(u) = \mu^{\prime} - \frac{2+s M_2}{2} u+ \frac{s M_4}{24} u^2\ .
\label{eq:momexpdispeg}
\end{equation}
The spatial eigenvalues are,
\begin{equation}
u_0=-\lambda_0^2=6 \frac{2+s M_2\pm \sqrt{(2+sM_2)^2-2\mu' s M_4/3}}{s M_4} .
\label{eq:momexeig}
\end{equation}
The RDZ manifold of $\tilde \Gamma (u)$, which signals HH and BD transitions, is given by $\tilde \Gamma(u_c)=\tilde \Gamma'(u_c)=0$:
\begin{eqnarray}
4 \mu^{\prime}-(2+s M_2) u_c=0 \label{eq:momexpdz1}\\
s M_4 u_c= 6(2+s M_2)\ .
\label{eq:momexpdz2}
\end{eqnarray}
Combining these two equations in order to eliminate $u_c$ yields the RDZ manifold,
\begin{equation}
\mu^{\prime}_{\rm RDZ} =\frac{3(2+s M_2)^2}{2s M_4} \ ,
\label{eq:momexpdzm}
\end{equation}
which is of codim-$1$ in the 3-D $(\mu^{\prime},sM_2,s M_4)$ parameter space. Since we are considering $sM_4<0$, $\mu'_{\rm RDZ}$ is always negative. Setting $u_c=0$ in (\ref{eq:momexpdz1}--\ref{eq:momexpdz2}) one obtains the QZ manifold
\begin{equation}
\mu^{\prime}_{\rm QZ}=0\, , \quad s_{\rm QZ} =-2/M_2 \ .
\label{eq:QZmomexp}
\end{equation}

For a fixed $sM_4$, considering the $(s,\mu^{\prime})$ parameter space the RDZ manifold has the shape of a parabola with vertex at the QZ point and unfolding towards negative $\mu^{\prime}$. In the part of the RDZ with $s<s_{\rm QZ}$, $u_c>0$, and corresponds to a MI while the other part corresponds to a BD. For parameter values in the region between the BD and MI lines, where the leading eigenvalues are a complex quartet, fronts connecting two equivalent homogeneous solutions has oscillatory tails and localized structures can be formed. 

In some cases one is interested in the effect of the nonlocality for fixed values of the parameters of the local GLE, namely for a given $\mu'$, then it is convenient to rewrite eq. (\ref{eq:momexpdzm}) so that $s$ is isolated
\begin{equation}
s_{\rm RDZ}=\frac{-(M_2+4\mu M_4)\pm 2\sqrt{4\mu^2 M_4^2+2\mu M_2 M_4}} {M_2^2}\ ,
\label{eq:scmomexp}
\end{equation}
where the $+$ solution corresponds to the BD transition and the $-$ solution to the MI.

For kernels whose moments can be written as $M_j=\sigma^j {\cal M}_j$ (cf. subsection IIIA of Part I), Eq.~(\ref{eq:scmomexp}) becomes,
\begin{equation}
s_{\rm RDZ}=\frac{1}{{\cal M}_2} \left[-\frac{1}{\sigma^2}-4\mu \frac{{\cal M}_4}{{\cal M}_2}\pm 2 \sqrt{4\mu^2 \frac{{\cal M}_4^2}{{\cal M}_2^2}+2\mu \frac{{\cal M}_4}{{\cal M}_2 \sigma^2} } \right] \, .
\label{eq:scmomexpsig}
\end{equation}
In the limit of nonlocal interaction range going to zero, $\sigma\rightarrow 0$, one has $s_{\rm RDZ}\rightarrow -1/({\cal M}_2\sigma^2)$ for both BD and MI transitions. For infinite range nonlocality, $\sigma\rightarrow\infty$, $s_{\rm RDZ}\rightarrow 0$ for the BD transition while $s_{\rm RDZ}\rightarrow -8\mu{\cal M}_4/{\cal M}_2^2$ for the MI. These predictions will be compared with the results for a Gaussian kernel in the next section.

\section{The Gaussian kernel} \label{Sect::gauss}

In this section we analyze the influence of a nonlocal Gaussian kernel in the shape of the front starting (or ending) at
an HSS of the GLE. Without loss of generality this kernel can be normalized so that $M_0=1$.
In terms of $u=-\lambda^2$ the dispersion relation obtained linearizing around the HSS can be written as (cf. Table~\ref{tab:kernels}),
\begin{equation}
\tilde{\Gamma}(u)=\mu^{\prime}-s-u+s \exp(-\sigma^2 u/2)\ .
\label{eq:gammagauss}
\end{equation}
The spatial eigenvalues are the zeros of (\ref{eq:gammagauss}), 
a transcendental equation that can be solved analytically in terms of the 
Lambert's W function (see Appendix~\ref{sec:Lambertap}).
The result is,
\begin{equation}
u_0=-\lambda_0^2=\mu^{\prime}-s+\frac{2}{\sigma^2} W_l\left[\frac{s\sigma^2}{2}
\exp\left(\frac{\sigma^2}{2} (-\mu^{\prime}+s) \right)\right]  \, ,
\label{eq:lambertsol}
\end{equation}
where $l\in  \mathbb{Z}$ and $W_l(x)$ is the $l$th branch of Lambert's W function,
$W(x)\in\mathbb{C}$, and, thus, the spectrum of spatial eigenvalues is infinite (numerable).

\begin{figure}
 \includegraphics[width=8.cm]{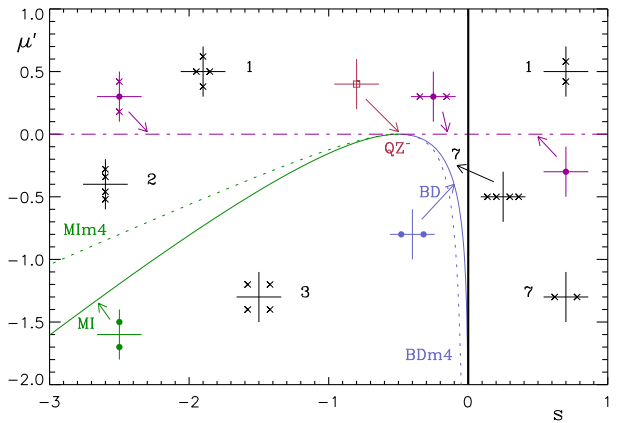}
 \caption{(Color online) Boundaries in the $(s,\mu')$ plane at which the leading spatial eigenvalues of the GLE with a Gaussian kernel exhibit different transitions for $\sigma=2$. Sketches indicate the location of the leading eigenvalues in the (${\rm Re}(\lambda),{\rm Im}(\lambda)$) plane ($\times$ signal simple eigenvalues, $\bullet$ double eigenvalues, and $\square$ quadruple eigenvalues). Dotted lines labeled as MIm4 and BDm4 show the MI and BD transitions given by the 4th moment expansion (\ref{eq:momexpdzm}).}
 \label{Fig::Gauss_muRDZ}
\end{figure}

In order to determine the location of the MI and BD instabilities of the HSSs, we look for the
RDZ of (\ref{eq:gammagauss}) which is given by $\tilde \Gamma(u_c)=\tilde \Gamma'(u_c)=0$:
\begin{eqnarray}
\exp(-\sigma^2 u_c/2)=-\frac{2}{s \sigma^2} \label{eq:k2temdgau}\\
\mu^{\prime}=s+\frac{2}{\sigma^2}+u_c\ .
\label{eq:muMI}
\end{eqnarray}
One consequence of (\ref{eq:k2temdgau}) is that BD and MI transitions require that $s<0$,
as $u_c$ has to be real. 
Combining (\ref{eq:k2temdgau}) and (\ref{eq:muMI}) to eliminate $u_c$ leads to the condition defining the RDZ manifold
of $\tilde{\Gamma}(u)$. This manifold has one dimension less than the dimensionality of the parameter space. Since we have three parameters, $\mu'$, $s$ and $\sigma$, the RDZ manifold is a 2-dimensional manifold given by
\begin{equation}
\frac{s\sigma^2}{2} \exp\left(\frac{\sigma^2}{2} (-\mu^{\prime}+s)\right)=-\frac{1}{e}\ .
\label{eq:sdefeqngk}
\end{equation}
The RDZ manifold given by (\ref{eq:sdefeqngk}) corresponds in Eq. (\ref{eq:lambertsol}) to the branching point
of the two real branches of Lambert's W function (see Appendix~\ref{sec:Lambertap}), in which these 
two branches both merge and finish. Solving Eq.~(\ref{eq:sdefeqngk}) for $\mu^{\prime}=\mu^{\prime}_{\rm RDZ}(s,\sigma)$ one has
\begin{equation}
 \mu^{\prime}_{\rm RDZ}=s+\frac{2}{\sigma^2}\left[1+\ln \left(-\frac{s\sigma^2}{2}\right)\right]\ .
\label{Eq::Gauss_muRDZ}
\end{equation}
A cut of the $\mu'_{\rm RDZ}$ manifold for $\sigma=2$ is shown in solid lines in Fig.~\ref{Fig::Gauss_muRDZ}.

The codim-$2$ QZ manifold (which is a line in our $3$-D parameter space), can be obtained by setting $u_c=0$ in (\ref{eq:k2temdgau}) and (\ref{eq:muMI}) or, alternatively, locating the sub-manifold of RDZ (\ref{eq:sdefeqngk}) in which $u_c=0$. The result is that the QZ manifold is defined by,
\begin{equation}
\mu^{\prime}_{\rm QZ}=0\, , \quad s_{\rm QZ} =-2/\sigma^2 \ .
\label{eq:QZgau}
\end{equation}
For the parameters of the QZ, $\tilde \Gamma''(0) <0$, therefore, in the notation of Part I, this is a QZ$^-$ point.  
In the $(s,\mu^{\prime})$ plane shown in Fig.~\ref{Fig::Gauss_muRDZ} the QZ point is located at (-1/2,0).
The character of the two pieces of the RDZ manifold, BD or MI, can be elucidated by calculating
$u_c$ on top of the manifold, such that, respectively, $u_c<0$ and $u_c>0$. Substituting (\ref{Eq::Gauss_muRDZ}) in (\ref{eq:muMI}) one obtains 
\begin{equation}
 u_c=\frac{2}{\sigma^2} \ln \left(-\frac{s\sigma^2}{2}\right)\ .
 \label{Eq::Gauss_uc}
\end{equation}
Therefore the part of the RDZ manifold in which $s\sigma^2<-2$ has a positive $u_c$ and corresponds to a MI while the part in which $0>s\sigma^2>-2$ corresponds to a BD. In Fig.~\ref{Fig::Gauss_muRDZ} the MI is located at the left of the QZ point and the BD at the right. Fronts starting (or ending) at the HSS have oscillatory tails for parameter values in the region between the MI and BD lines. This region is labeled as 3 in  Fig.~\ref{Fig::Gauss_muRDZ} in agreement with the notation used in Part I. The other parameter regions of the figure are also labeled as in Part I. For $s<0$ we refer to Part I for a detailed description of the regions and the transitions between them. 
The $s=0$ line corresponds to the GLE with local coupling for which there are only 2 spatial eigenvalues which are a real doublet for $\mu'<0$ and an imaginary doublet for $\mu'>0$. At $\mu=0$ the two components of the doublet collide at the origin (Hamiltonian-pitchfork bifurcation). For $s>0$, despite the presence of the Gaussian nonlocal kernel, the spatial dynamics shows a qualitative behavior is similar to that for $s=0$.

The second derivative of $\tilde \Gamma(u)$, given by
\begin{equation}
 \tilde \Gamma''(u)= \frac{s\sigma^4}{4} \exp(-u\sigma^2/2) \,
 \label{Eq::Gauss_Gamma_second}
\end{equation}
does not vanish for any value of $u$ provided $s \neq 0$. This indicates that there is no cusp point for the GLE with a Gaussian nonlocal kernel. For $s=0$ the second derivative vanishes but this corresponds to the GLE with only local interaction whose dispersion relation is linear in $u$, thus it has nothing to do with a cusp point.

We now look for 3DZ and 3DZ$(i\omega)$ codim-2 points which, as discussed in Part I, correspond to the coincidence of a simple zero at the origin $\tilde \Gamma(0)=0$ and a RDZ at finite distance, $\tilde \Gamma(u_c)=\tilde \Gamma'(u_c)=0$. The first condition, $\tilde \Gamma(0)=0$ implies $\mu=0$, thus 3DZ and 3DZ$(i\omega)$ can be obtained setting $\mu_{\rm RDZ}=0$ in Eq. (\ref{Eq::Gauss_muRDZ}) and looking for solutions with non-zero $u_c$. There is no such a solution and therefore the GLE with a Gaussian nonlocal kernel does not have any 3DZ$(i\omega)$ or 3DZ points. 

The absence of cusp and 3DZ points, indicates that the GLE with a Gaussian nonlocal kernel does not have any crossover manifold. This has strong implications on the location of the complex quartets in the $({\rm Re}(\lambda), {\rm Im}(\lambda))$ plane. In particular if a real doublet is leading the spatial dynamics, changing parameters complex quartets can not overcome the real doublet. In this case, the only way oscillatory tails can appear is after a BD transition in which two real doublets collide to become a complex quartet.

\begin{figure}[t!]
\begin{center}
\includegraphics[width=8.cm]{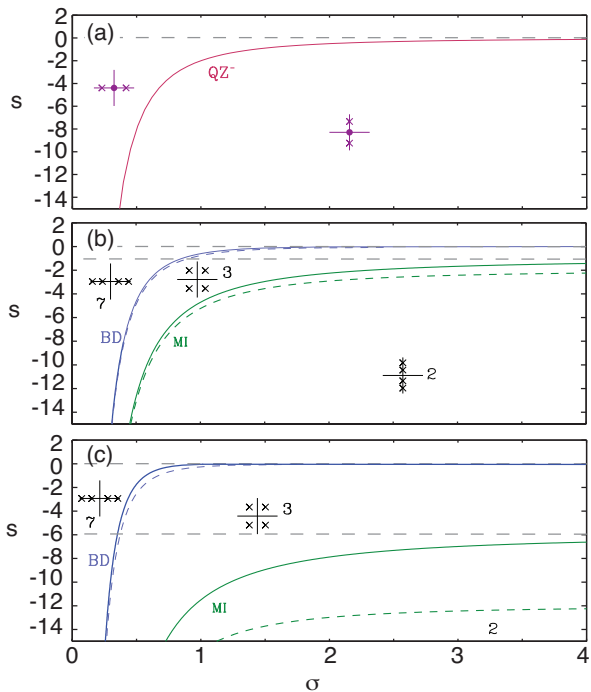}
\caption{(Color online) Boundaries in the $(\sigma,s)$ plane at which the leading spatial eigenvalues of the GLE with a Gaussian kernel exhibit different transitions. Panel (a) shows the QZ manifold at $\mu'=0$. Panels (b) and (c) show the BD and MI manifolds at $\mu'=-1$ and $\mu'=-6$ respectively. The short-dashed curves show the BD transition as predicted by the 4th moment expansion, Eq.~(\ref{eq:scmomexpsig}). Long-dashed horizontal lines show the asymptotic values for $\sigma \rightarrow \infty$. Sketches represent the location of the leading eigenvalue.}
\label{Fig::SpatialEigBoundary_Sigma_S_realGLE}
\end{center}
\end{figure}

We now consider the effect of the nonlocal Gaussian kernel for given values of the parameters of the local dynamics, namely for a given $\mu^{\prime}$. Solving (\ref{eq:sdefeqngk}), e.g. for $s_{\rm RDZ}(\mu^{\prime},\sigma)$, one obtains,
\begin{equation}
s_{\rm RDZ}=\frac{2}{\sigma^2} W_l\left[-\exp\left(\frac{\sigma^2}{2} \mu^{\prime}-1\right)\right]\ ,
\label{eq:ssigmagauss}
\end{equation}
that leads to two real branches since the argument of $W$ is in the interval $[-1/e,0]$. These two pieces of the RDZ manifold are organized by the codim-$2$ QZ manifold (\ref{eq:QZgau}). 
Fig.~\ref{Fig::SpatialEigBoundary_Sigma_S_realGLE} shows three cuts of the $s_{\rm RDZ}(\mu^{\prime},\sigma)$ for different values of 
$\mu^{\prime}$. At $\mu^{\prime}=0$ one has the QZ manifold (\ref{eq:QZgau}) [Fig.~\ref{Fig::SpatialEigBoundary_Sigma_S_realGLE} (a)], from which the BD and MI branches emerge as $\mu^{\prime}$ is decreased [see panel (b)]. The upper branch has $u_c<0$ and therefore it corresponds to a BD while the lower branch corresponds to the MI. LSs exist for parameter values in the region between the BD and MI curves. The BD and MI branches separate as $\mu^{\prime}$ is further decreased [see panel (c)], thus the region where LSs exists becomes larger. The asymptotic limit of both transitions as $\sigma\rightarrow\infty$ is $s_{\rm BD}(\sigma\rightarrow\infty)\rightarrow 0^-$ and
$s_{\rm MI}(\sigma\rightarrow\infty)\rightarrow \mu^{\prime}$ 
\footnote{This can be shown taking into account that for the argument of $W$ in (\ref{eq:ssigmagauss}), 
$\lim_{\sigma\rightarrow\infty} -\exp\left(\frac{\sigma^2}{2} \mu^{\prime}-1\right)\rightarrow 0^-$,
$z(\sigma\rightarrow \infty)\rightarrow 0^-$, 
and remembering (see Appendix~\ref{sec:Lambertap}) that $W_0(0^-)\rightarrow 0$, 
and so $s_{\rm BD}\rightarrow 0^-$ while
as $W_{-1}(0^-)\rightarrow -\infty$ it is necessary to use its series expansion. Keeping the first term,
$W_{-1}(z)\sim \ln(-z)$ one gets $s_{MI}\rightarrow 2/\sigma^2 ((\sigma^2\mu^{\prime}/2-1))\sim \mu^{\prime}$,
as shown in Fig.~\ref{Fig::SpatialEigBoundary_Sigma_S_realGLE}.}.

\begin{figure}[t!]
\begin{center}
\includegraphics[width=8cm]{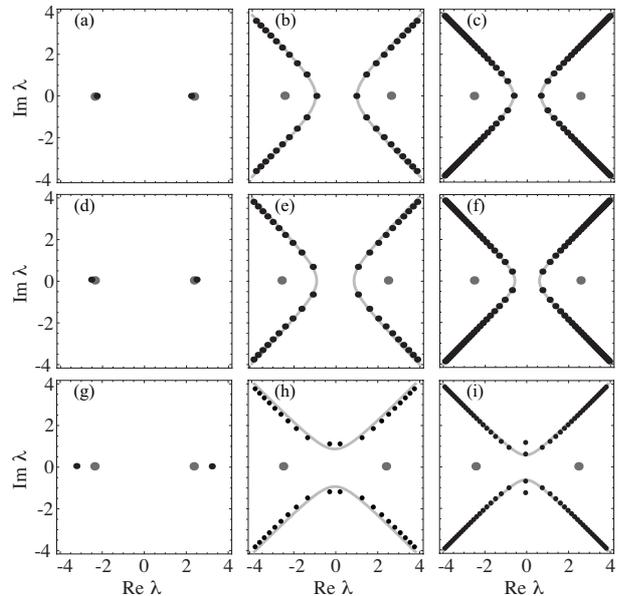}
\caption{Location of the first spatial eigenvalues for the GLE with a Gaussian nonlocal kernel in the complex $\lambda$ plane (shown as black dots) for $\mu'=-6$. For comparison the two grey dots show the location of the eigenvalues for the local GLE ($s=0$). The hyperbola, given by Eq.~(\ref{Eq::spatialDyn_realGLE4_amp}), in greyscale represents an approximation for the location of the spatial eigenvalues. For panels (a-c) on the top row, $s=1$, for (d-f) on the middle row $s=-1$ and for (g-i) on the bottom row $s=-7.5$.
For panels (a),(d) and (g) on the left column $\sigma=0.3$, for (b), (e) and (h) on the middle column $\sigma=2$ and
for (c), (f) and (i) on the right column $\sigma=3$. For all values of $\sigma$ the number of spatial eigenvalues is infinite: the plot just presents the region around the origin in the complex plane.}
\label{Fig::spatEig_hyperbola}
\end{center}
\end{figure}

\begin{figure}
\begin{center}
\includegraphics[width=8.6cm]{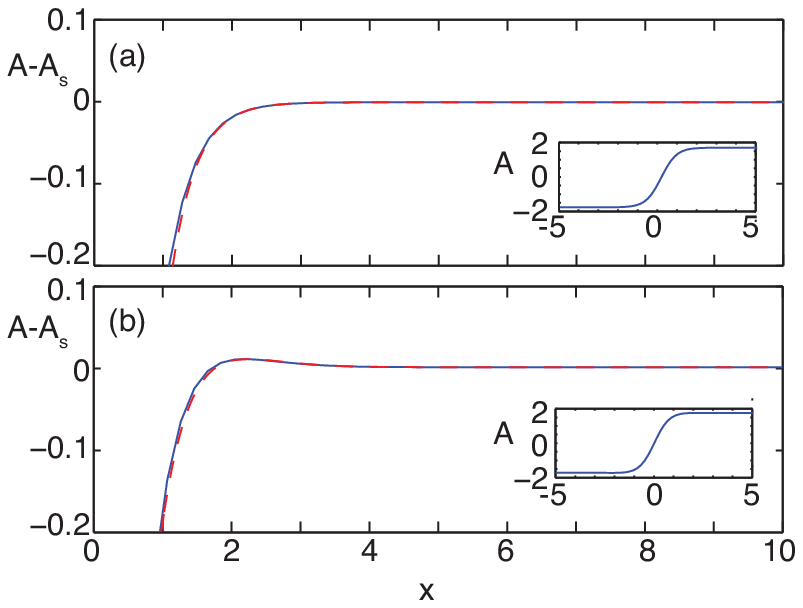}
 \caption{(Color online) The solid line shows front profile for the GLE with a Gaussian nonlocal kernel for $\mu'=-6$, $s=-1$ and (a) $\sigma=0.5$ and (b) $\sigma=1$. The red dashed lines show the approximations given by Eqs.~(\ref{Eq::Front_fit_monotonous}) and (\ref{Eq::Front_fit_BD}) (see text).
\label{Fig::Gauss_front_profile}
}
\end{center}
\end{figure}

To illustrate the main behaviors exhibited by the system we plot in Fig.~\ref{Fig::spatEig_hyperbola}
the location in the complex $\lambda$ plane of the first few spatial eigenvalues of (\ref{eq:lambertsol}) 
for $\mu^{\prime}=-6$ and three values of $s$ (at different rows) and three values $\sigma$ (at different columns).
For attractive nonlocal interaction, $s>0$, only the principal branch, $W_0$, is real thus the spectrum contains only one real doublet [panels (a-c)]. For small $\sigma$ it is located very close to the real doublet of the local dynamics, as shown in panel (a). There is also an infinite number of complex eigenvalues but they are located outside the region shown in panel (a). As $\sigma$ increases the location of the spatial eigenvalues approaches the imaginary axis. Since, as discussed before, the GLE with Gaussian kernel has no crossover manifolds, the real doublet is always the eigenvalue located closer to the imaginary axis (see panels (b) and (c)). Neither a BD transition can exist for $s>0$ because there is no other real doublet with which the leading real doublet can collide.
As a consequence for $s>0$ the spatial dynamics is always lead by a real doublet and fronts decay monotonically. 

For repulsive nonlocal interaction, $s<0$, and $\sigma$ small, the argument of $W_l(x)$ in (\ref{eq:lambertsol}) is in the range $x\in [-1/e,0]$, and both $W_0(x)$ and $W_{-1}(x)$ are real (see Appendix~\ref{sec:Lambertap}) and, as a consequence, there are two pairs of real eigenvalues. The pair located closer to the origin is shown in panel (d) for $s=-1$ and in panel (g) for $s=-7.5$. Increasing $\sigma$ the two real doublets approach each other and collide at the BD transition, which corresponds to the branching point of $W$ beyond which there is no real solution. Panels (e) and (h) correspond to parameters at the right of the BD curve in Fig.~\ref{Fig::SpatialEigBoundary_Sigma_S_realGLE}), and one finds a leading complex quartet. Fig.~\ref{Fig::Gauss_front_profile} illustrates the change of the front profile when crossing the BD line. From Eq.~(\ref{eq:sdefeqngk}) one obtains that for $\mu'=-6$ and $s=-1$ the BD line is located at $\sigma=\sqrt{(2/5)W_0(5/e)} \approx 0.5708076$. For $\sigma=0.5$, at the left of the BD line, the fronts are monotonic. Fig.~\ref{Fig::Gauss_front_profile} (a) shows the detailed shape of the front close to the HSS corresponding to $A_s=\sqrt{\mu}$. The overall profile of the front connecting the two HSS is shown in the inset. Close to the HSS the front is well described by an exponential of the form 
\begin{equation}
 A(x)-A_{st} \approx c_1 e^{q_1 x}\ ,
 \label{Eq::Front_fit_monotonous}
\end{equation}
where $q_1=-2.753$ is the leading spatial eigenvalue and the coefficient $c_1$ has been fitted to $c_1=-4.507$. When crossing the BD line oscillations in the front profile appear initially with an infinite wavelength. The front profile for $\sigma=1$ is shown in Fig.~\ref{Fig::Gauss_front_profile} (b). Again close to the HSS the front profile is very well described by an exponential of the form
\begin{equation}
 A(x)-A_{st} \approx c_1 e^{q_1 x} \cos(k_1 x + \phi_1) 
  \label{Eq::Front_fit_BD}
\end{equation}
where $q_1=-2.01$ and $k_1=1.01$ are the real and imaginary part of the spatial eigenvalue and the coefficients $c_1=2.02$ and $\phi_1=2.91$ have been fitted. 

Once crossed the BD line, for $s=-1$ increasing $\sigma$ the spatial eigenvalues get closer to the imaginary axis. Nevertheless for $s>\mu'$ the situation remains qualitatively the same no matter how large is $\sigma$, as shown in Fig.~\ref{Fig::spatEig_hyperbola} (f). For $s<\mu'$, as the range of interaction $\sigma$ increases one crosses the MI line, so that the HSS becomes modulationally unstable. Beyond the MI line the spatial dynamics is lead by two imaginary doublets as shown in  Fig.~\ref{Fig::spatEig_hyperbola} (i).

The spatial eigenvalues lie on a hyperbola-like curve for $\sigma$ high enough (cf. Fig.~\ref{Fig::spatEig_hyperbola}). 
Although the analytical solution is available, (\ref{eq:lambertsol}), it does
not yield a geometrically transparent picture of the locus of the curve on which the spatial
eigenvalues lie. 
This behavior can be easily understood by neglecting the linear term versus
the exponential one in (\ref{eq:gammagauss}),
\begin{equation}
\exp (-\sigma^2 u_0/2) = 1-\mu'/s\ .
\label{Eq::spatialDyn_realGLE2old}  
\end{equation}
Using  $u_0=-(q_0+ik_0)^2$ one gets
\begin{equation}
\exp \left[\sigma(q_0^2-k_0^2)/2 + i \sigma q_0 k_0 \right] = 1-\mu'/s\ .
\label{Eq::spatialDyn_realGLE2B}  
\end{equation}
From the modulus of (\ref{Eq::spatialDyn_realGLE2B}) one has,
\begin{equation}
 q_0^2-k_0^2 = \frac{2}{\sigma^2} \log\left|1-\frac{\mu'}{s}\right| , \label{Eq::spatialDyn_realGLE4_amp}
\end{equation}
which represents a hyperbola with eccentricity $\sqrt{2}$ in the complex plane. 
The RHS of (\ref{Eq::spatialDyn_realGLE2B}) is real and for $\mu'/s<1$ is positive, thus the
phase of the exponent must be 0 or multiple of $2\pi$: 
\begin{equation}
 q_0 k_0 = \frac{2 n \pi}{\sigma^2} \, , \quad n \in {\cal Z} \, .
 \label{Eq::spatialDyn_realGLE4}
\end{equation}
Conversely for $\mu'/s>1$, the RHS of (\ref{Eq::spatialDyn_realGLE2B}) is negative and
\begin{equation}
 q_0 k_0 = \frac{(2 n + 1) \pi}{\sigma^2} \, , \quad n \in {\cal Z} \, .
 \label{Eq::spatialDyn_realGLE5}
\end{equation}
Eqs. (\ref{Eq::spatialDyn_realGLE4}) and (\ref{Eq::spatialDyn_realGLE5}) can be seen as a 
selection criterion, which has to be satisfied by a point on the hyperbola to be
a spatial eigenvalue of the system. These hyperbolas are shown in Fig.~\ref{Fig::spatEig_hyperbola}.
The approximation is meaningless for $\sigma \ll 1$ (cf. Fig.~\ref{Fig::spatEig_hyperbola}(a), (d) and (g)).
When $\sigma \geq 2$ it is clear that the above equation of the hyperbola 
provides a good approximation of the location of the spatial eigenvalues and the 'selection criterion' in fact gives eigenvalues that lie increasingly close to the real ones. Notice that for panels in top and middle rows $\mu'/s>1$, thus the selection criterion is given by (\ref{Eq::spatialDyn_realGLE5}). Instead for the panels in the bottom row $\mu'/s<1$, thus the criterion is given by (\ref{Eq::spatialDyn_realGLE4}). Therefore the hyperbola in the panels of the bottom row has a conjugated shape as compared to the one in the panels of the top and middle rows.

\begin{figure}[t!]
\begin{center}
\includegraphics[width=8.6cm]{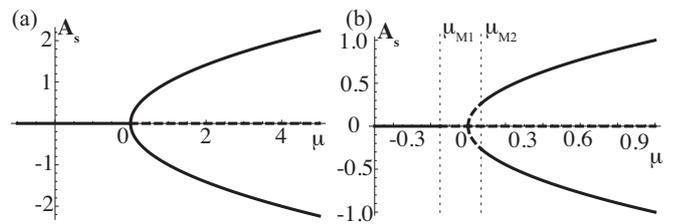}
\caption{Bifurcation diagram of the HSS of the nonlocal GLE with a Gaussian kernel and for $\sigma = 2$: (a) $s = 0$ (local case); (b) $s = -1$. Stable solutions are shown in a solid line, while the unstable ones in a dashed line.}
\label{Fig::realGLE_hombif}
\end{center}
\end{figure}

\begin{figure}[t!]
\begin{center}
\includegraphics[width=8.6cm]{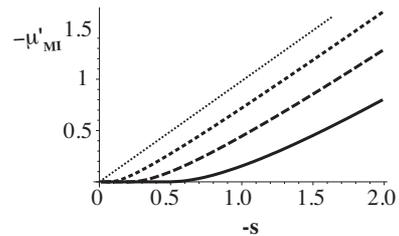}
\caption{Location of the MI in parameter space for the GLE with Gaussian kernel as a function of the nonlocal strength $s$ for different values of the interaction range: The solid, long-dashed and short-dashed lines correspond, respectively, to $\sigma = 2, 3, 5$, while the dotted line represents the $\sigma\rightarrow\infty$ limit.
}
\label{Fig::MI_GLE}
\end{center}
\end{figure}

Regarding temporal instabilities associated to the MI, there is a finite range of values of $\mu\in [\mu_{M1},\mu_{M2}]$ around
the pitchfork bifurcation, $\mu^{\prime}=0$, where HSSs are modulationally unstable. For the 
parameters of Fig.~\ref{Fig::realGLE_hombif}, $\sigma=2$ and $s=-1<-1/(2\sigma^2)$, 
$\mu^{\prime}_{\rm MI}=-1+(1+\ln 2)/2=-0.153426$, and, thus, $\mu_{\rm M1}=\mu^{\prime}_{\rm MI}$ and
$\mu_{\rm M2}=-\mu^{\prime}_{\rm MI}/2=-0.076713$.  In turn,
in Fig.~\ref{Fig::MI_GLE} the dependence of $\mu'_{\rm M1}$ on
$s$ for three different values of $\sigma$ is plotted. In the limit of $\sigma\rightarrow\infty$
this dependence is given by the line $\mu^{\prime}=s$.
The effect of a nonlocal nonlinear response in a MI is also
discussed in Ref.\ \cite{Krolikowski_PRE_2001, *Wyller_PRE_2002}.

Finally, we compare the results obtained here with an expansion up to the 4th moment as discussed in Sect.~\ref{Sect::moments}. Fig. 
\ref{Fig::Gauss_muRDZ} shows in dashed lines the location of the BD and MI manifolds given by (\ref{Fig::Gauss_muRDZ}). The prediction given by the 4th moment expansion is quite good for the BD transition. The prediction for the MI manifold, while following the correct trend, becomes quite off as soon as one moves away from the QZ point where $u_c=0$.
As for the dependence of $s_{\rm RDZ}$ on the kernel width for a fixed $\mu'$, one can use Eq.~(\ref{eq:scmomexpsig}) with ${\cal M}_2=1$ and ${\cal M}_4=3$ (see Table \ref{tab:kernels}). As shown in  Fig.~\ref{Fig::SpatialEigBoundary_Sigma_S_realGLE} for the BD transition, the result given by 4th moment approximation (dashed line) is in good agreement with the exact one (solid line). In the case of the MI, the 4th order expansion does work as well. In the limit of zero interaction range, it correctly predicts an asymptotic behavior $s_{\rm RDZ}\rightarrow -1/\sigma^2$, however in the limit of infinite interaction range, $\sigma\rightarrow\infty$, the prediction is $s_{\rm RDZ}\rightarrow -24\mu$ which is out of the figure. Therefore the results for the MI given by the 4th order expansion are outside the parameter region plotted in Fig.~\ref{Fig::SpatialEigBoundary_Sigma_S_realGLE}.

\section{The mod-exponential kernel}\label{Sect::expon}

We consider here the effect of a kernel whose profile decays exponentially in space on the tails of fronts starting (or ending) in an HSS of the GLE. In Fourier space the mod-exponential kernel is a Lorentzian (cf. Table \ref{tab:kernels})
\begin{equation}
\tilde{\hat{K}}(u)= \frac{1}{1+4\sigma^2 u}\ ,
\label{Eq::Lorentzian_u}
\end{equation} 
which in the complex plain has a singularity at $u=-1/(4 \sigma^2)$. The dispersion relation obtained linearizing around the HSS is given by,  
\begin{equation}
\tilde \Gamma(u)=\mu^{\prime}-u-s+ \frac{s}{1+4\sigma^2 u}\ .
\label{eq:dispersion_exp_kernel_u}
\end{equation}
The spatial eigenvalues are the zeros of $\tilde \Gamma(u)$. In this case there are only 4 spatial eigenvalues $\lambda_0$ given by
\begin{align}
u_0 & =-\lambda_0^2= \nonumber \\
   & =\frac{1}{2} \left[\mu'-s-\frac{1}{4\sigma^2}\pm\sqrt{\left(s-\mu'+\frac{1}{4\sigma^2}\right)^2+\frac{\mu'}{\sigma^2}}
\right].
\label{eq:speigenexp}
\end{align}

MI and BD instabilities are located on the RDZ manifold of $\Gamma(u)$ which is given by $\tilde \Gamma(u_c)=\tilde\Gamma'(u_c)=0$:
\begin{eqnarray}
u_c&=&\frac{-1\pm 2\sigma\sqrt{-s}}{4\sigma^2} \label{eq:k2temdexp}\\
\mu^{\prime}&=&s+\frac{1}{4\sigma^2}+2 u_c\ .
\label{eq:muMIexp}
\end{eqnarray}
MI and BD transitions require $u_c$ real, thus MI and BD transitions can only exist for $s<0$, namely, for repulsive nonlocal interaction. 
Combining (\ref{eq:k2temdexp}) and (\ref{eq:muMIexp}) to eliminate $u_c$ one obtains the RDZ manifold which in the $(\mu',s,\sigma)$ parameter space is the surface given by
\begin{equation}
\mu^{\prime}_{\rm RDZ}=s-\frac{1}{4\sigma^2}\pm\frac{\sqrt{-s}}{\sigma}\ .
\label{Eq::Exponential_muRDZ}
\end{equation}

Setting $u_c=0$ in (\ref{eq:k2temdexp}) and (\ref{eq:muMIexp}) one obtains the QZ codim-2 bifurcations (lines in the 3D parameter space). It turns out that there are two QZ lines. The first one takes place for finite $\sigma$ and is given by
\begin{equation}
 \mu'_{\rm QZ1}=0, \, \, s_{\rm QZ1}=-\frac{1}{4 \sigma^2} \ .
 \label{eq:quadzero_exponential}
\end{equation}
For the parameters of QZ1, $\tilde \Gamma''(0)=32 s_{\rm QZ1} \sigma^4=-8 \sigma^2<0$, thus, in the notation of Part I, this is a QZ$^-$ point. The second QZ line is located at
\begin{equation}
 \sigma_{\rm QZ2}=\infty, \quad \mu'_{\rm QZ2}=s \, .
\end{equation}
For the parameters of QZ2, $\tilde \Gamma''(0)= 32 s \sigma_{\rm QZ2}^4 =-\infty$, thus this is also a QZ$^-$ point, albeit a particular one since the second derivative is infinite. As a consequence the parabola described by the RDZ close to QZ2 is infinitely narrow, and the BD and MI lines unfold from QZ2 practically tangentially.

In the part of the RDZ manifold that starts from the side $s \sigma^2 < -1/4$ of QZ1 and the side $s<\mu'$ of QZ2, $u_c>0$, it thus corresponds to a MI bifurcation. On the other part $u_c<0$, thus corresponds to a BD.

Cusp or 3DZ codim-2 points can not exist for the GLE with a mod-exponential kernel since they require at least 6 spatial eigenvalues. The crossover manifold does not exist either in this case.

\begin{figure}[t!]
\begin{center}
\includegraphics[width=8cm]{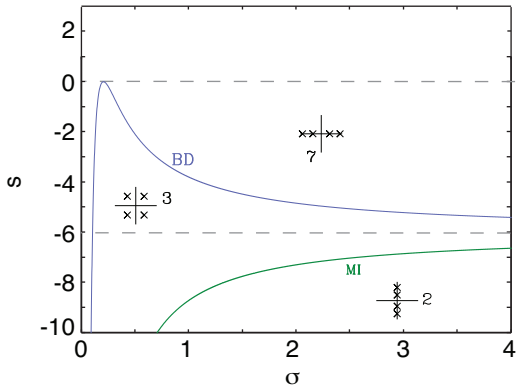}
\caption{\label{Fig::s_sigma_nlRGLE_Laplacian_mu_3} 
(Color online) Boundaries in the $(\sigma, s)$-plane separating the regions of monotonic and oscillatory tails in the GLE with a mod-exponential kernel for $\mu'=-6$ obtained from (\ref{eq:ssigmaexp}). The horizontal dashed line at $s=-6$ shows the asymptotic limit of MI and BD lines for $\sigma \rightarrow \infty$.
}
\end{center}
\end{figure}

We now address the effect of the mod-exponential kernel for a given value of $\mu'$. It is convenient to rewrite (\ref{Eq::Exponential_muRDZ}) as,
\begin{equation}
s_{\rm RDZ}=\mu^{\prime}-\frac{1}{4\sigma^2}\pm \frac{1}{\sigma} \sqrt{-\mu^{\prime}}\ .	
\label{eq:ssigmaexp}
\end{equation}
The $+$ and $-$ signs correspond to the BD and MI manifolds respectively. 
The MI and BD transitions are shown in Fig.~\ref{Fig::s_sigma_nlRGLE_Laplacian_mu_3} for the non-zero HSSs for $\mu'=-6$ corresponding to $\mu=3$. LSs are found in the parameter region bounded by the BD and MI curves. The existence of QZ2 leads to a significant difference with the Gaussian kernel (cf. Fig.~\ref{Fig::SpatialEigBoundary_Sigma_S_realGLE}). Increasing $\sigma$ the BD and MI lines tend asymptotically one to the other and meet at the QZ2. As a consequence the region of LSs narrows as $\sigma$ increases. The curve $s_{\rm BD}$ has a maximum at $\sigma^{\dag}=1/(2\sqrt{2\mu})$ (i.e., at $\sigma=1/\sqrt{24}=0.204124$ in Fig.~\ref{Fig::s_sigma_nlRGLE_Laplacian_mu_3}, where it reaches $s=0$, the maximum value of $s$ for which the RDZ manifold exists. 
For $\mu\rightarrow 0$ the BD and MI curves approach each other and meet at QZ1.

Considering the MI and comparing with the local case shown in Fig.~\ref{Fig::realGLE_hombif}(a), one also finds that there is a finite range of values of $\mu\in[\mu_{M1},\mu_{M2}]$ around the pitchfork bifurcation, $\mu=0$, that are modulationally unstable. For the parameters of Fig.~\ref{Fig::realGLE_hombif}(b), $\sigma=2$ and $s=-1<-1/(4\sigma^2)$,
$\mu_{M1}=-9/16$ and $\mu_{M2}=-\mu_{M1}/2=9/32$.

Written in terms of $u$ the mod-exponential kernel in Fourier space (\ref{Eq::Lorentzian_u}) has a pole of order 1 at $u=-1/(4 \sigma^2)$. As a consequence a moment expansion around $u=0$ such as the one discussed in section 3.1 of Part I will converge only for $|u|<1/(4 \sigma^2)$, and therefore it will be of limited use. A truncation up to $M_4$ identifies the spatial behavior around the first QZ, in particular the existence of spatial tails, and thus of LSs, for  $s<0$ but as $\sigma$ is increased the predictions from moment expansion are quite off. This kernel can be considered as the simplest with singularities and no further approximations can be obtained from a Laurent expansion (Sect. 3.2 of Part I) since $1/(1+4\sigma^2u)$ is already the first and only term of that expansion. Still, proceeding as in Sect. 3.2 of Part I, one can obtain an exact transformation for the nonlocal interaction term. Using (\ref{Eq::Lorentzian_u}) the nonlocal interaction can be written as
\begin{equation}
  \hat{F}(k,\sigma) = \frac{1}{1+4\sigma k^2} \hat{A}(k)\ ,
\end{equation}
or equivalently
\begin{equation}
 (1+4\sigma k^2) \hat{F}(k,\sigma) = \hat{A}(k)\ .
\end{equation}
In real space this leads to
\begin{equation}
 (1-4\sigma \partial_{xx})F(x,\sigma) = A(x)\ .
\end{equation}
which is an ordinary differential equation. Therefore the GLE with a mod-exponential nonlocal kernel can be written as a partial differential equation coupled to an ordinary differential equation,
\begin{eqnarray}
&&\partial_t A  =(\mu - s) A - A^3 + \partial_{xx} A + s  F(x,\sigma) \nonumber \\
&&\partial_{xx} F(x,\sigma) = \frac{1}{4\sigma} (F(x,\sigma)-A(x)) \, ,
\label{eq:GLE_exponential_kernel_transform}
\end{eqnarray}
where we have used that $M_0=1$. This treatment of the mod-exponential kernel was introduced in Ref. \cite{Ermentrout93}, and used also by other authors \cite{Coombes05,Clerc_PRE_2010}.

\section{The Mexican-hat kernel}\label{Sect::Mexhat}

In this Section, we discuss in detail the effects of a spatially nonlocal kernel that is not everywhere positively defined.
More precisely the kernel we consider consists of two Gaussians and has an extra parameter $b>0$ (cf. Table \ref{tab:kernels}) 
that regulates the spatial extension of the negative sector of the kernel. The total area of this kernel is given by $M_0=2(1-b)$ (cf. Table \ref{tab:kernels}), and for $b>1$, $M_0<0$. 
For $s>0$  one has short-range attraction (activation) and medium to long-range repulsion (inhibition). In this case, activation dominates globally for $b<1$ while otherwise overall inhibition is stronger than activation.
For $s>0$  one has inhibition in the short-range and activation in the medium to long-range and globally inhibition dominates for $b<1$ while activation does otherwise.

The dispersion relation for this kernel is (cf. Eq. (\ref{Eq::HSS_stab_u}) and Table~\ref{tab:kernels}), 
\begin{equation}
\tilde{\Gamma}(u)=\mu^{\prime}- 2s (1 - b) - u + 2 s ( 1 - b + b \sigma^2 u ) e^{-\sigma^2 u/2},
\label{Eq::Mexhat_Gamma_u}
\end{equation}
and setting $\tilde{\Gamma}(u)=0$ does not lead to a closed expression for the spatial eigenvalues, that
have to be obtained numerically for this kernel. The RDZ manifold is given by $\tilde \Gamma(u_c)=\tilde \Gamma'(u_c)=0$:
\begin{align}
\frac{1}{s\sigma^2}&=(-1+3b-b\sigma^2 u_c) e^{-\sigma^2 u_c/2}
\label{Eq::Mexhat_RDZ1}\\
\mu^{\prime}&=2s-2sb + \frac{2}{\sigma^2} + u_c - 4 s b e^{-\sigma^2 u_c/2} .
\label{Eq::Mexhat_RDZ2}
\end{align}
Since now the parameter space  $(\mu,s,\sigma,b)$ is $4$-D, the RDZ manifold is a 3-D hyper-surface.
Eq. (\ref{Eq::Mexhat_RDZ1}) can be solved analytically for $u_c$,
\begin{equation}
u_{c,l}= \frac{1}{\sigma^2} \left[ 3 - \frac{1}{b} -2 W_l(\chi)\right]\ .
\label{Eq::Mexhat_uc}   
\end{equation}
where $l=0,-1$ are the indices of the two real branches of the Lambert $W$ function (see Appendix \ref{sec:Lambertap}) and 
\begin{equation}
 \chi=\chi(b,s,\sigma)=\frac{1}{2bs\sigma^2}\exp\left(\frac{3b-1}{2b}\right)\ .
\end{equation}
Substituting (\ref{Eq::Mexhat_uc}) into (\ref{Eq::Mexhat_RDZ2}) one gets
\begin{equation}
\mu^{\prime}_{{\rm RDZ,}l}=2s(1-b) + \frac{5}{\sigma^2} -\frac{1}{b\sigma^2} -\frac{2W_l(\chi)}{\sigma^2} -\frac{2}{\sigma^2 W_l(\chi)} \ .
 \label{eq:mexhat_mu_rdz}
\end{equation}
For $-1/e < \chi < 0$ the RDZ manifold has two branches which we label $l=0$ and $l=-1$ as the indices of the $W$ function (see Appendix \ref{sec:Lambertap}). For $\chi > 0$ the RDZ manifold has a single branch given by $l=0$. For $\chi < -1/e$, $W$ does not take real values thus there is no RDZ manifold. The asymptotic behavior of the RDZ branches for large $s$ is given by
\begin{align}
 \mu'_{{\rm RDZ,}0}=&2s\left[1-b-2b \exp{\left(\frac{1-3b}{2b}\right)}\right]+\frac{3b-1}{b\sigma^2}  \nonumber \\ & +\mathcal{O}(s^{-1}) \label{Eq::Mexhat_muRDZ0_asymptotic} \\
 \mu'_{{\rm RDZ},-1} =& 2s(1-b)+  \mathcal{O}\left(\log(s)\right) \label{Eq::Mexhat_muRDZ-1_asymptotic} \, .
\end{align}

The part of the RDZ manifold with $u_c<0$ corresponds to a BD transition while the part with $u_c>0$ corresponds to a HH bifurcation. Note that the BD and HH parts of the RDZ manifold are not directly related to the index $l$ of $\mu^{\prime}_{{\rm RDZ,}l}$. Instead, as it will be discussed below, $\mu^{\prime}_{{\rm RDZ,}0}$ have both BD and HH parts and the same applies to $\mu^{\prime}_{{\rm RDZ,}-1}$. For the moment, we will distinguish the part of the BD and MI manifolds in which the second derivative of the dispersion relation $\Gamma(u)$, given by
\begin{equation}
 \tilde \Gamma''(u)=\frac{s\sigma^4}{2}\left(1-5b+b\sigma^2 u\right) e^{-\sigma^2 u/2},
 \label{Eq::Mexhat_Gamma_second}
\end{equation}
is positive from that where is negative. The part of the RDZ with $\tilde \Gamma''(u_c)>0$ corresponds to a local minimum of the dispersion relation crossing zero and in this section it will be referred to as HH$^+$ or BD$^+$, while the part with $\tilde \Gamma''(u_c)<0$ corresponds to a local maximum of the dispersion relation crossing zero and will be referred to as as HH$^-$ or BD$^-$. If the local maximum signaled by the HH$^-$ turns out to be the global maximum then it corresponds to a modulational instability of the HSS and, as in previous sections, it will be referred to as MI.   

\begin{figure*}
\includegraphics{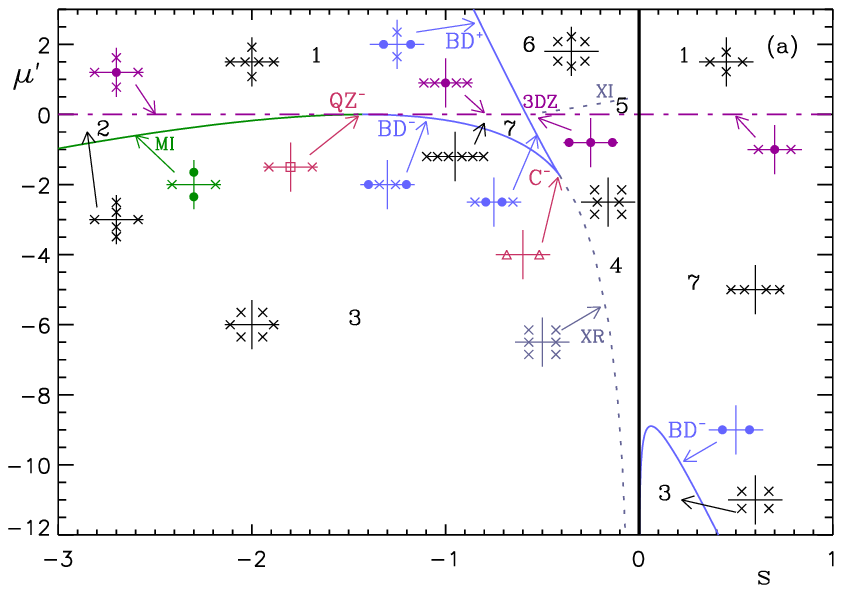}
\includegraphics{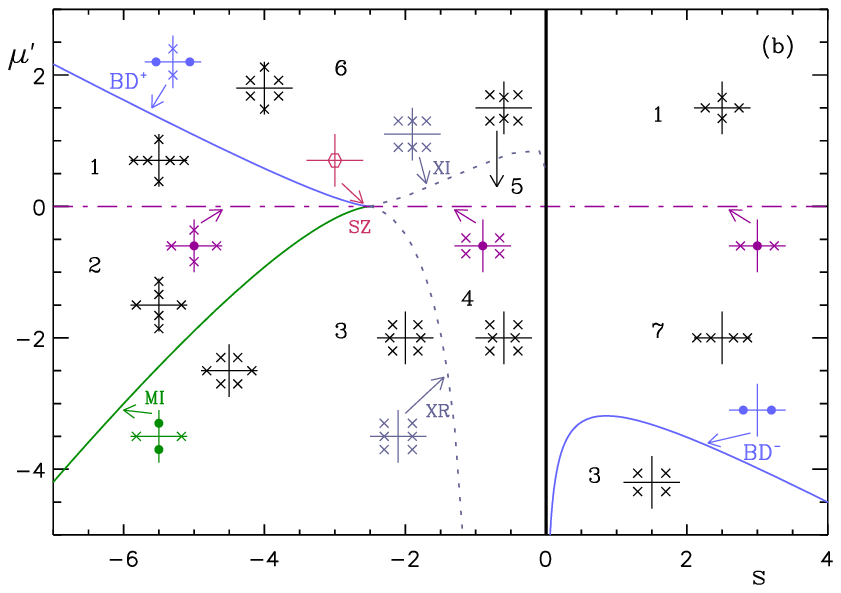}
\includegraphics{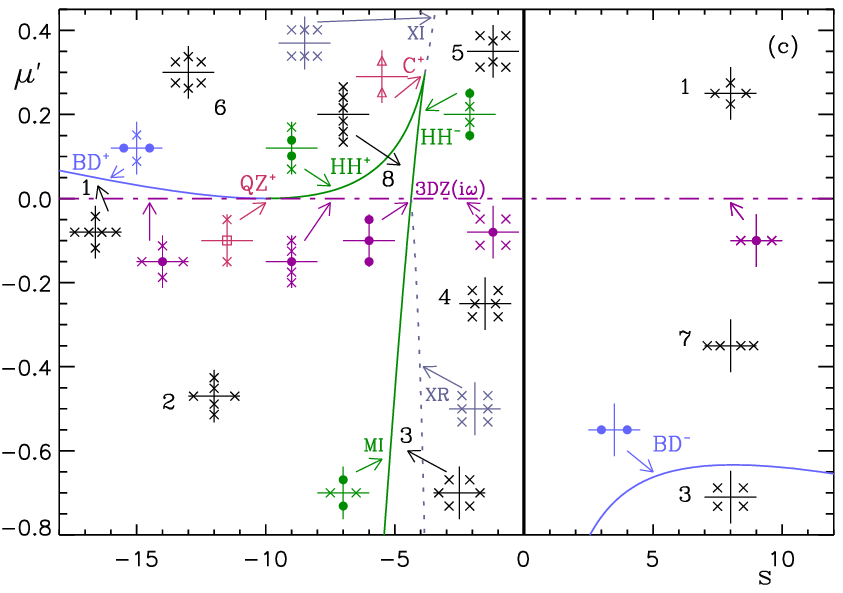}
\includegraphics{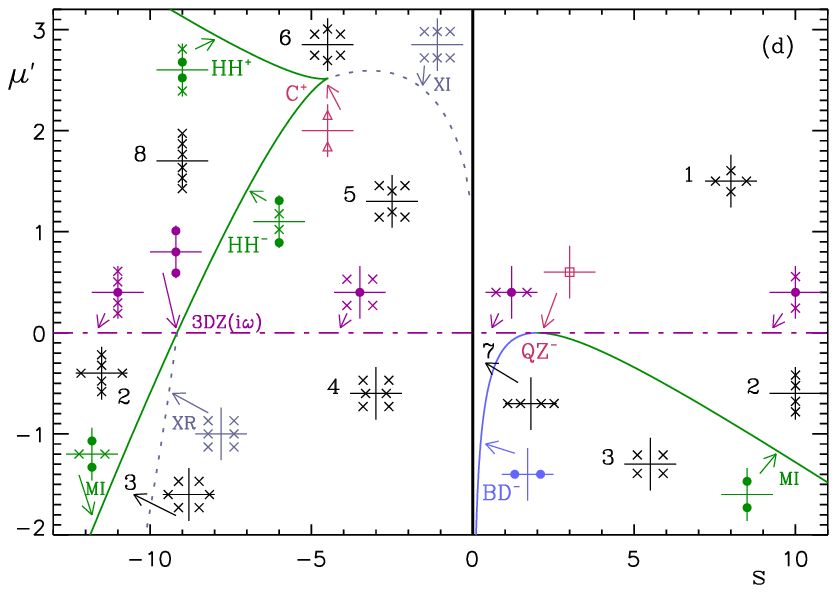}
\caption{(Color online) Boundaries in the $(\mu',s)$ parameter space at which the leading spatial eigenvalues of the GLE with a Mexican-hat nonlocal kernel exhibit different transitions for $\sigma=1$ and (a) $b=0.1$, (b) $b=b_{\rm SZ}=0.2$, (c) $b=0.3$ and (d) $b=0.5$. Sketches indicate the position of the zeros of $\Gamma_s(\lambda)$ in the (${\rm Re}(\lambda),{\rm Im}(\lambda)$) plane ($\times$ signal simple zeros, $\bullet$ double zeros, $\triangle$ triple zeros, $\square$ quadruple zeros, and $\hexagon$ sextuple zeros).}
\label{Fig::Mexhat_muRDZ}
\end{figure*}

Setting $u_c=0$ in Eqs. (\ref{Eq::Mexhat_RDZ1}--\ref{Eq::Mexhat_RDZ2}) one finds the QZ manifold, which
in the $(\mu,s,\sigma,b)$ parameter space is a 2-D surface given by,
\begin{equation}
\mu_{\rm QZ}^{\prime}=0 \, ; \quad s_{\rm QZ}=\frac{1}{\sigma^2 (3b-1)}\ .
\label{Eq::Mexhat_QZ}
\end{equation}
$s_{\rm QZ}$ has a divergence at $b=1/3$, and, as a consequence, the QZ manifold splits in two parts. For $b<1/3$, $s_{\rm QZ}<0$ as shown in Fig.~\ref{Fig::Mexhat_muRDZ} (a), while for $b>1/3$, $s_{\rm QZ}>0$ as shown in Fig.~\ref{Fig::Mexhat_muRDZ} (d). This is a clear distinction with the previous two kernels for which $s_{\rm QZ}$ was always negative. 

As discussed in Part I there are two kinds of QZ points depending on the sign of $\Gamma''(u)$ at the QZ point. From (\ref{Eq::Mexhat_Gamma_second}) we have
\begin{equation}
\tilde \Gamma_{\rm QZ}'' (0) =\frac {s_{\rm QZ} \sigma^4}{2}(1-5b)=-\frac{1}{\sigma^2}\frac{1-5b}{1-3b}
\label{Eq::Mexhat_Gamma_second_QZ}
\end{equation}
For $\tilde \Gamma_{\rm QZ}''(0)<0$ the QZ is a QZ$^-$ unfolding a BD$^-$ and a MI manifolds towards $\mu'<0$ [see Fig.~\ref{Fig::Mexhat_muRDZ} (a) or (d)], while for $\tilde \Gamma_{\rm QZ}''(0)>0$ one has a QZ$^+$ unfolding a BD$^+$ and a HH$^+$ manifolds towards $\mu'>0$ [see Fig.~\ref{Fig::Mexhat_muRDZ} (c)]

At difference with the previous kernels, the GLE with a Mexican-hat nonlocal kernel
exhibits a codim-$2$ cusp manifold. Setting $\tilde \Gamma''(u_{\rm cusp})=0$ one has,
\begin{equation}
 u_{\rm cusp}=\frac{1}{\sigma^2}\left(5-\frac{1}{b}\right)\, .
 \label{Eq::Mexhat_u_cusp}
\end{equation}
From $\tilde \Gamma'(u_{\rm cusp})=0$ one obtains,
\begin{equation}
 s_{\rm cusp}=-\frac{1}{2b\sigma^2}\exp\left(\frac{5}{2}-\frac{1}{2b}\right)\, .
 \label{Eq::Mexhat_s_cusp}
\end{equation}
Finally setting $\tilde \Gamma(u_{\rm cusp})=0$ and using (\ref{Eq::Mexhat_s_cusp}) one arrives to,
\begin{equation}
 \mu^{\prime}_{\rm cusp}=\frac{1}{\sigma^2} \left[9-\frac{1}{b}+\left(1- \frac{1}{b}\right) \exp\left(\frac{5}{2}-\frac{1}{2b}\right)\right]\, .
 \label{Eq::Mexhat_mu_cusp}
\end{equation}

From Eq.~(\ref{Eq::Mexhat_u_cusp}) and Eq.~(\ref{Eq::Mexhat_uc}), one finds that at the cusp $W_l(\chi_{\rm cusp})=-1$, which is the branching point of the Lambert $W$ function where the two real branches $W_0$ and $W_{-1}$ originate (cf. Appendix~\ref{sec:Lambertap}). These two $W$ branches associated to the two branches of the $\mu^{\prime}_{\rm RDZ}$ manifold which in parameter space emerge from the cusp one tangent to the other [see for example Fig.~\ref{Fig::Mexhat_muRDZ} (a)]. Since $W_{-1}(\chi)<-1$ for any value of $\chi$ (cf. Appendix~\ref{sec:Lambertap}), on the branch $\mu^{\prime}_{{\rm RDZ},-1}$, $u_{c,1}>u_{\rm cusp}$ and, as a consequence, the sign of the second derivative $\tilde \Gamma''(u_{c,1})$ is that of $s$. 
On the contrary on the branch $\mu^{\prime}_{{\rm RDZ},0}$, $u_{c,0}<u_{\rm cusp}$ and $\tilde \Gamma''(u_{c,0})>0$ for $s<0$ while $\Gamma''(u_{c,0})<0$ for $s>0$.

For $u_{\rm cusp}<0$ the cusp is, in the notation of Part I, a C$^-$ unfolding a BD$^+$ and a BD$^-$ manifolds. If the cusp is located at $s<0$ as is the case of Fig.~\ref{Fig::Mexhat_muRDZ} (a), then the BD$^+$ corresponds to $\mu^{\prime}_{{\rm RDZ},0}$ and the BD$^-$ to  $\mu^{\prime}_{{\rm RDZ},-1}$. Moving away from the cusp along the BD$^-$ line, $u_{c,-1}$ increases and eventually it reaches zero at the QZ$^-$ where the BD$^-$ becomes a MI. The C$^-$ cusp also unfolds a crossover manifold at which a real doublet and a complex quartet are located at the same distance from the imaginary axis which we label as XR. In parameter space XR can be seen as the natural continuation of the two BD manifolds that end one tangent to the other at the cusp [see Fig.~\ref{Fig::Mexhat_muRDZ} (a)]. 

For $u_{\rm cusp}>0$ the cusp is a C$^+$ unfolding a HH$^+$ and a HH$^-$ which for a cusp located at $s<0$ correspond to $\mu^{\prime}_{{\rm RDZ},0}$ and to $\mu^{\prime}_{{\rm RDZ},-1}$, respectively [see Fig.~\ref{Fig::Mexhat_muRDZ} (c)]. Moving along the HH$^+$ manifold away from the cusp, $u_{c,0}$ decreases and eventually it reaches zero at the QZ$^+$ point where the HH$^+$ becomes a BD$^+$. Also unfolding from C$^+$ there is a crossover manifold (labeled as XI) at which a imaginary doublet and a complex quartet are located at the same distance from the real axis. Similarly as before, in parameter space XI can be seen as the continuation of the two HH manifolds ending at the cusp. 

The GLE with a Mexican-hat kernel also has a 3DZ and a 3DZ$(i\omega)$ codim-2 points. As discussed in part I, at these points a simple zero at the origin $\tilde \Gamma(0)=0$ and a RDZ at finite distance, $\tilde \Gamma(u_c)=\tilde \Gamma'(u_c)=0$, take place simultaneously. Since $\tilde \Gamma(0)=0$ implies $\mu=0$, the 3DZ and 3DZ$(i\omega)$ can be obtained setting $\mu_{\rm RDZ}=0$ in Eq. (\ref{eq:mexhat_mu_rdz}) and looking for solutions with non-zero $u_c$. Thus, within the $\mu=0$ hyper-plane the location of the 3DZ and 3DZ$(i\omega)$ is given by the implicit equation
\begin{equation}
 0=2s\sigma^2(1-b) + 5 -\frac{1}{b} -2W_l(\chi) -\frac{2}{W_l(\chi)}\ .
 \label{Eq::Mexhat_3DZ}
\end{equation}
The 3DZ$(i\omega)$ point is located on the HH$-$ manifold that unfolds from the C$^+$ cusp [see Figs.~\ref{Fig::Mexhat_muRDZ} (c) and (d)] and have a significant effect on it. At this point HH$-$, which at the cusp is a local maximum of the dispersion relation crossing zero, becomes a global maximum. Thus the HH bifurcation becomes a MI [see Fig. 8 of Part I]. Similarly the 3DZ point is located on the BD$^+$ manifold that unfolds from the C$^-$ cusp [see Fig.~\ref{Fig::Mexhat_muRDZ} (a)].

The sextuple zero point, which organizes the overall scenario, takes place when $u_{\rm cusp}=0$, which in the $(\mu,s,\sigma,b)$ parameter space is the line
\begin{equation}
 b_{\rm SZ}=\frac{1}{5}\, ; \quad s_{\rm SZ}=-\frac{5}{2\sigma^2}\, ; \quad  \mu_{\rm SZ}=0\ .
 \label{eq::Mexhat_SZ}
\end{equation} 

The parameter space portrait for the GLE with a Mexican-hat kernel is as follows. 
For $b=b_{\rm SZ}$ [Fig.~\ref{Fig::Mexhat_muRDZ} (b)] the SZ unfolds a BD$^+$ ($\mu^{\prime}_{{\rm RDZ},0}$), a MI  ($\mu^{\prime}_{{\rm RDZ},-1}$), a XR and a XI manifolds in a similar way as described in Part I for the six-order dispersion relation in $\lambda$. For $s$ negative and large the asymptotic behavior of BD$^+$ is given by Eqs.~(\ref{Eq::Mexhat_muRDZ0_asymptotic}) and that of MI by (\ref{Eq::Mexhat_muRDZ-1_asymptotic}).
The part for $s<0$ of Fig.~\ref{Fig::Mexhat_muRDZ} (b) can be directly compared with Fig. 1(b) of Part I. We refer to Part I for a detailed explanation of all the regions surrounding the SZ as well as the transitions between them. The regions relevant for the existence of stable LSs are region 3 where the spatial dynamics is lead by a complex quartet and the part of region 4 close to the crossover XR where the spatial dynamics results from the combination of a real doublet and a complex quartet. The $s=0$ line corresponds to the GLE with local coupling with two spatial eigenvalues which are real $\mu'<0$ and imaginary for $\mu'>0$. For $s>0$ and $\mu<0$ there is a BD$^-$ line given by $\mu^{\prime}_{{\rm RDZ},0}$ which separates region 3 led by a complex quartet from region 7 led by two real doublets, in which fronts are monotonic.  For large positive $s$ 
follows the asymptotic behavior (\ref{Eq::Mexhat_muRDZ0_asymptotic}). Thus the BD$^+$ line for $s<0$ and the BD$^-$ line for $s<0$ have the same oblique asymptote. For $s \rightarrow 0$ the BD$^-$ line goes to $-\infty$. When crossing $\mu=0$ from region 7 into 1, the components of the doublet closer to the origin collide leading to a imaginary doublet. This is a Hamiltonian-pitchfork bifurcation as described in Part I.

For $b<b_{\rm SZ}$ the SZ unfolds a QZ$^-$ within the $\mu'=0$ hyperplane and a cusp located at $\mu'<0$ as shown in Fig.~\ref{Fig::Mexhat_muRDZ} (a) for $b=0.1$. From Eq.~(\ref{Eq::Mexhat_s_cusp}), for $b<b_{\rm SZ}$, $u_{\rm cusp} <0$, thus this is a C$^-$ cusp unfolding two BD manifolds and a crossover XR. The BD$-$ manifold (given by $\mu^{\prime}_{{\rm RDZ},-1}$) connects C$^-$ with QZ$^-$ where it becomes an MI. The BD$+$ manifold (given by $\mu^{\prime}_{{\rm RDZ},0}$) connects C$^-$ with the 3DZ located at the $\mu'=0$. For large negative $s$ the asymptotic behavior of both manifolds is given by Eqs.~(\ref{Eq::Mexhat_muRDZ0_asymptotic}) and (\ref{Eq::Mexhat_muRDZ-1_asymptotic}) respectively.
The overall picture shown in the part part of Fig.~\ref{Fig::Mexhat_muRDZ} (a) corresponding to $s<0$ has the same structure as Fig. 1(a) of Part I, to which we refer for a discussion. As for $s>0$ the regions are the same as in Fig.~\ref{Fig::Mexhat_muRDZ} (b) but the BD$^-$ line given by $\mu^{\prime}_{{\rm RDZ},0}$ has moved down to more negative values of $\mu'$ [Notice the different vertical scale in Figs.~\ref{Fig::Mexhat_muRDZ} (a) and (b)] and region 3 has narrowed. As a consequence one has a large parameter region for $s<0$ where localized structures may be formed which includes region 3 unfolding from QZ$^-$ and the part of region 4 close to XR while for $s>0$ there is a narrow region 3 located at small $s$ and reachable only for strongly negative values of $\mu'$.

If $b$ is further decreased the $C^-$ cusp moves towards smaller values of $s$ and more negative values for $\mu'$. At the same time the BD$^+$ line born at the right of the cusp becomes more vertical. The BD$^-$ line at $s>0$ is located further down and region 3 keeps narrowing. In the limit $b \rightarrow 0^+$ the cusp goes to $s_{\rm cusp} \rightarrow 0^+$ and $\mu'_{\rm cusp}\rightarrow -\infty$ while the BD$^-$ line located in the $s>0$ semi-plane goes also to $-\infty$. The result is that one recovers the parameter space diagram obtained for the Gaussian kernel (Fig.~\ref{Fig::Gauss_muRDZ}). 

For $b>b_{\rm SZ}$ the SZ unfolds a QZ$^+$ and a 3DZ$(i\omega)$ manifolds located within the $\mu'=0$ hyperplane and a cusp located at $\mu'>0$ as shown in Fig.~\ref{Fig::Mexhat_muRDZ} (c) for $b=0.3$. From Eq.~(\ref{Eq::Mexhat_s_cusp}), for $b_{\rm SZ}$, $u_{\rm cusp} >0$, thus this is a C$^+$ cusp unfolding two HH manifolds and a crossover XI. The HH$^+$ manifold (given by $\mu^{\prime}_{{\rm RDZ},0}$) connects C$^+$ with QZ$^+$ where it becomes a BD$^+$. The HH$^-$ manifold (given by $\mu^{\prime}_{{\rm RDZ},-1}$) connects C$^+$ with the 3DZ$(i\omega)$. Globally the part for $s<0$ of Fig.~\ref{Fig::Mexhat_muRDZ} (c) has the same regions and transitions as those obtained for $b>0$ for the six-order dispersion relation considered in Part I (Fig.~1 (c)). For $s<0$ region 3 now unfolds from the 3DZ$(i\omega)$ point and has a sharp-pointed shape and thus is narrower than in Figs.~\ref{Fig::Mexhat_muRDZ} (a) and (b).
For  $s>0$ the regions are the same as in Fig.~\ref{Fig::Mexhat_muRDZ} (b) but the BD$^-$ line given by $\mu^{\prime}_{{\rm RDZ},0}$ has moved up and region 3 has widened significantly. Still in order to reach region 3 for $s>0$ it is necessary that $\mu'$ is not too close to zero.

If $b$ is further increased the QZ$^+$ point moves towards more negative values for $s$. For $b \rightarrow 1/3^-$, $s_{\rm QZ} \rightarrow -\infty$ [cf. Eq.~(\ref{Eq::Mexhat_QZ})]. Also the slope and the ordinate at the origin of the oblique asymptote of $\mu^{\prime}_{{\rm RDZ},0}$ tend to zero as $b \rightarrow 1/3$ as can be seen from Eq.~(\ref{Eq::Mexhat_muRDZ0_asymptotic}). 
For $b>1/3$, the QZ point is located at $s_{\rm QZ}>0$ starting from $s_{\rm QZ}=\infty$ at $b=1/3^+$ and monotonically approaching $s_{\rm QZ}=0$ as $b$ increases. The QZ is in fact a QZ$^-$ point since for $b>1/3$, $\tilde \Gamma''_{\rm QZ}(0)<0$ [see Fig.~\ref{Fig::Mexhat_muRDZ} (d) for $b=1/2$]. The QZ$^-$ unfolds a BD$^-$ and a MI manifolds given by $\mu^{\prime}_{{\rm RDZ},0}$. The cusp C$^+$ is still present in the $\mu'>'$, $s<0$ quadrant, unfolding a HH$+$ line which asymptotically connects with the MI line for $s>0$, since both are part of $\mu^{\prime}_{{\rm RDZ},0}$. As before the HH$-$ unfolding from C$^+$ goes to the 3DZ$(i\omega)$ point where it becomes an MI. Now one has region 3 reaching all the way up to $\mu^{\prime}=0$ for both $s<0$ and $s>0$, and, in fact, region $3$ is much larger for $s>0$ than for $s<0$. From a physical point of view this can be understood by noticing that for $b>1/3$ all the moments of the Mexican-hat kernel, except $M_0$ are negative. Thus while the overall area of 
the kernel is positive and for $s>0$ the nonlocal interaction can be considered as globally attractive, at medium and long distances the nonlocal interaction is repulsive. This compensates the local attractive interaction leading to oscillations in the front profile.  

Increasing $b$ in the half-plane $s>0$ as $b$ increases the QZ$^-$ approaches zero but there are no qualitative changes.
In the half-plane $s<0$ the slope of the HH$^-$ that unfolds from C$^+$ given by by $\mu^{\prime}_{{\rm RDZ},-1}$ decreases [e.g. Eq.~(\ref{Eq::Mexhat_muRDZ-1_asymptotic})] and the 3DZ$(i\omega)$ point moves towards more negative values of $s$. For $b=1$ the 3DZ$(i\omega)$ point is located at $s->-\infty$. For $b>1$ the HH$^-$ does not cross $\mu'=0$ and there is no 3DZ$(i\omega)$ point and, thus, no crossover XR nor region 3 for $s<0$. The result is that for $b>1$ only region 3 unfolded by QZ$^-$ located at $s>0$ remains as parameter regions where stable LSs can be formed.
 
We now focus on the effect of the Mexican-hat nonlocal kernel for given parameter values of the local GLE, that is for a given $\mu^{\prime}$. In what follows we take $b = 1/2$ [Fig.~\ref{Fig::Mexhat_muRDZ} (d)] so that the total area of the kernel is $M_0=1$, as in the kernels considered in the previous Sections. 

\begin{figure}[t!]
\begin{center}
\includegraphics[width=8cm]{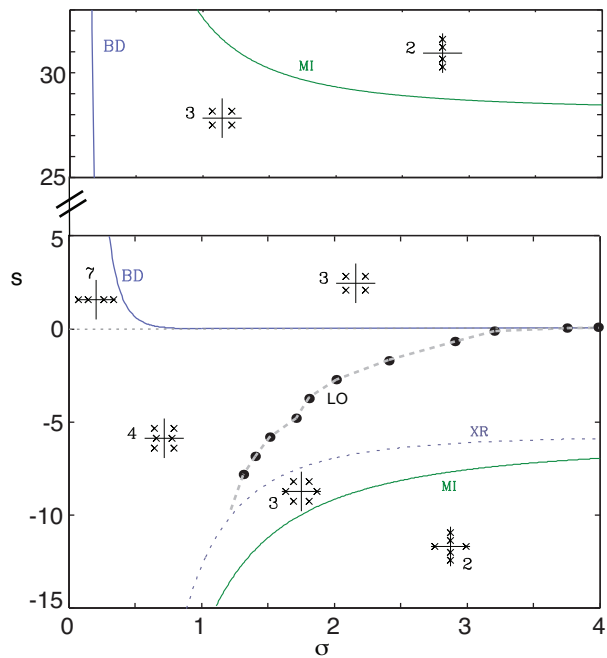}
\caption{\label{Fig::mexHat_bound} (Color online) Boundaries in the $(\sigma, s)$-plane separating the regions of monotonic 
and oscillatory tails when using a Mexican-hat shaped kernel in the nonlocal GLE. $\mu^{\prime} = -6$.}
\end{center}
\end{figure}

For this kernel a closed formula for $s_{\rm RDZ}(\mu^{\prime},\sigma)$ is not available, but, nevertheless, it can be found semi-analytically by replacing (\ref{Eq::Mexhat_uc}) into (\ref{Eq::Mexhat_RDZ2}) with $b = 1/2$. The result for $\mu'=-6$ is shown in Fig.~\ref{Fig::mexHat_bound}. As discussed above, the result is that one finds two sections of the RDZ manifold for $s>0$, BD$^-$ and MI, unfolding from QZ$^-$ while for $s<0$ one has a MI and a crossover XR unfolding from 3DZ$(i\omega)$ [see also Fig.~\ref{Fig::Mexhat_muRDZ} (d)]. 

\begin{figure}[t!]
\begin{center}
\includegraphics[width=8cm]{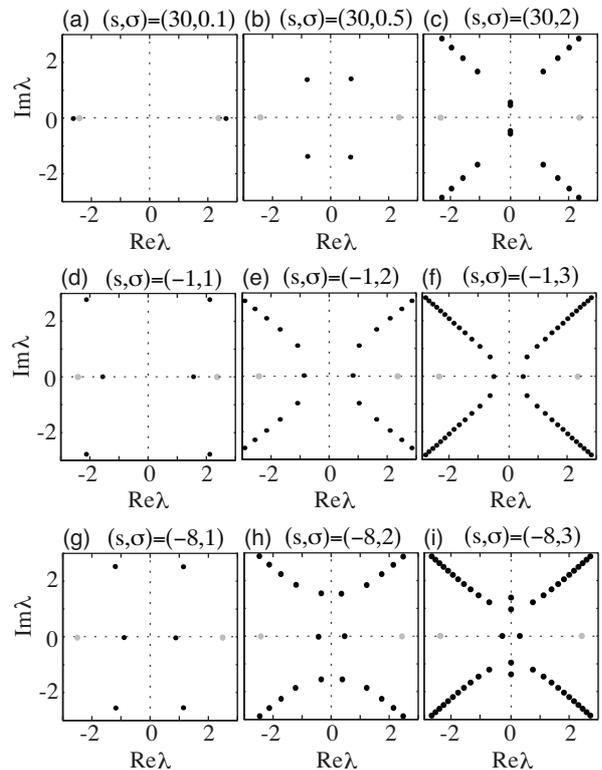}
\caption{\label{Fig::mexHatspatEig} Location of the first spatial eigenvalues for the GLE with a Mexican-hat nonlocal kernel in the complex $\lambda$ plane (shown as black dots) for $b=1/2$ and $\mu^{\prime}=-6$. The top row corresponds to $s=30$ with (a) $\sigma=0.1$, (b) $\sigma=0.5$, and (c) $\sigma=2$. The middle row corresponds to $s=-1$ with (d) $\sigma=1$, (e) $\sigma=2$,
and (f) $\sigma=3$. The bottom row corresponds to $s=-8$ with (g) $\sigma=1$, (h) $\sigma=2$, and (i) $\sigma=3$.
For comparison the two grey dots show the location of the eigenvalues for the local GLE ($s=0$). In all the cases the number of spatial eigenvalues is infinite: the plot just presents the region around the origin in the complex plane.}
\end{center}
\end{figure}

Fig.~\ref{Fig::mexHatspatEig} shows the location in the complex $\lambda$ plane of the first few spatial eigenvalues for different values of $s$ and $\sigma$. The first row corresponds to $s=30$. For $\sigma$ small the spatial dynamics is lead by a real doublet located close to the real doublet of the GLE with local coupling [Fig.~\ref{Fig::mexHatspatEig} (a)]. There is another real doublet located outside the region plotted in Fig.~\ref{Fig::mexHatspatEig} (a). Increasing $\sigma$ the second real doublet gets closer to the origin and collides with the first pair in a BD transition leading to a complex quartet  [Fig.~\ref{Fig::mexHatspatEig} (b)] within parameter region 3, where front tails have a oscillatory profile. Further increasing $\sigma$ leads to a collision of the components of the complex quartet on the imaginary axis (MI transition) which results in two imaginary doublets Fig.~\ref{Fig::mexHatspatEig} (c)] within parameter region 2.

For $b=1/2$ and $s<0$ the one finds only one real doublet in the spatial spectrum. For $s$ not too negative this real doublet leads the spatial dynamics for any value of $\sigma$ as shown in the second row of Fig.~\ref{Fig::mexHatspatEig} which corresponds to $s=-1$.
The three panels of this row are within parameter region 4.

For a large negative $s$ the real doublet leads the dynamics only for small $\sigma$ [see Fig.~\ref{Fig::mexHatspatEig} (g)]. Increasing (g) one encounters the crossover XR after which there is a complex quartet located closer to the imaginary axis than the real doublet [Fig.~\ref{Fig::mexHatspatEig} (h)] and therefore one enters in the parameter region 3 unfolding from the 3DZ$(i\omega)$ point. For larger values of $\sigma$ there is a MI transition at which the components of the complex quartet collide on the imaginary axis. After this the spatial dynamics is lead by two imaginary pairs as shown in Fig.~\ref{Fig::mexHatspatEig} (i) for parameters within region 2.

\begin{figure}[t!]
\begin{center}
\includegraphics[width=8cm]{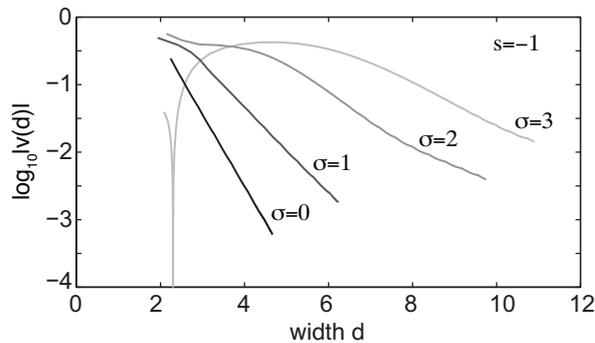}
\caption{\label{Fig::Mexhat_frontvelocity} Velocity at which two fronts connecting the two equivalent HSSs approach each other for the GLE equation with a Mexican-hat nonlocal kernel with $b=1/2$, $\mu'=6$, $s=-1$ and different values for the interaction range $\sigma$ within parameter region 4.
}
\end{center}
\end{figure}

As discussed in Part I in the part of region 4 located close to the crossover XR, the real part of the complex quartet is close to that of the real doublet and the spatial dynamics results, in fact, from the combination of the real doublet and the complex quartet. Thus, although asymptotically the front tail is monotonic, closer to the front core the complex quartet manifest introducing oscillations, whose role in the fronts interaction can lead to the existence of stable LSs. At difference with the case where the leading eigenvalues are complex, and the front tail has oscillations asymptotically, here the front shows oscillations only close to the core, and thus the locking of two fronts can only occur at short distances. 

To determine the part of region 4 where this occurs we consider the full evolution equation (\ref{Eq::RGL}) with the Mexican-hat nonlocal kernel. We set the initial condition such that there are two fronts connecting the two equivalent HSSs and look at the velocity at which the two fronts approach each other. In Ref. \cite{GelensPRL2010} (cf. Fig. 1) it was shown that the interaction of two monotonic fronts (in systems with two equivalent states) decays exponentially (both for local and spatially nonlocal interactions where the kernel decays faster than exponentially). In the part of region 4 where the fronts are no longer monotonic, the envelope of the interaction of two oscillatory fronts still decays exponentially, but at some particular distances the fronts pin and the relative velocity drops to zero. Fig.~\ref{Fig::Mexhat_frontvelocity} shows the dependence of the relative velocity $v(d)$ on the distance between the fronts $d$ for a interaction for different values of the interaction range $\sigma$.

\begin{figure}[t!]
\begin{center}
\includegraphics[width=8cm]{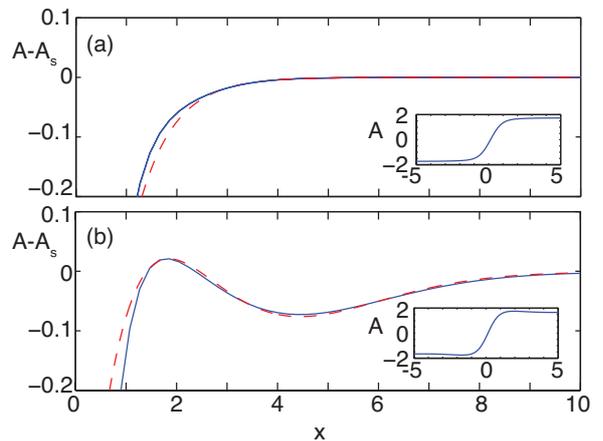}
\caption{(Color online) Spatial front profile for the GLE with a Mexican-hat nonlocal kernel for $b=1/2$, $\mu'=-6$, $s=-1$ and (a) $\sigma=1$ and (b) $\sigma=3$. The red dashed lines show the approximations given by Eqs.~(\ref{Eq::Front_fit_monotonous}) and (\ref{Eq::Front_fit_crossover}) (see text).
\label{Fig::Mexhat_profile}
}
\end{center}
\end{figure}

One can see that when the range of nonlocal interaction vanishes ($\sigma=0$) the logarithm of the relative velocity grows linearly when decreasing $d$. Switching on the nonlocal interaction with a small interaction range (e.g. $\sigma=1$) a similar linear growth is encountered, albeit with a smaller slope, for $d > 3$. For smaller values of $d$ the two fronts have a stronger interaction and the logarithm of the velocity increases linearly when decreasing $d$ but at a much slower rate. For $\sigma=1$ although the real part of the complex quartet is not far away from the real doublet, as shown in Fig.~\ref{Fig::mexHatspatEig} (d), the separation is still sufficient to warrant a quasi-monotonic front shape as shown in Fig.~\ref{Fig::Mexhat_profile} (a). Close to the HSS the front profile is well described by an exponential of the form (\ref{Eq::Front_fit_monotonous}) with $q_1=-1.433$, which is the value for leading spatial eigenvalue, and fitting the amplitude to $c_1=-1.299$. As a consequence, the velocity at which two connected fronts approach each other is a monotonic function of the front separation.

As the range of the nonlocal interaction $\sigma$ increases, the real part of the complex quartet keeps approaching the real doublet as shown in Fig. \ref{Fig::mexHatspatEig} (e) for $\sigma=2$, and a plateau appears in the velocity when the distance is around 3 (see line for $\sigma=2$ in Fig.~\ref{Fig::Mexhat_frontvelocity}). For $\sigma=3$ clearly the velocity goes to zero at $d=d_0\approx 2.3$. Fronts starting at a initial distance larger than this one will approach each other until reaching $d_0$ while fronts starting at a distance slightly smaller than $d_0$ will separate until reaching $d_0$. Thus $d_0$ is a stable distance at which the fronts lock forming a LS. 

Proceeding in this way one can determine numerically the boundary within region 4 where the oscillations induced by the complex quartet close to the front core enable the formation of LSs. This boundary is labeled as LO in Fig.~\ref{Fig::mexHat_bound}. For small nonlocal interaction ranges the line LO approaches to the crossover XR. This comes from the fact that for small values of $\sigma$ the real parts of the doublet and the quartet separate faster than for large values of $\sigma$, as illustrated in Fig.~\ref{Fig::mexHatspatEig} (compare panels (d) and (g) for instance). Thus the parameter region where one must account for both the real doublet and the complex quartet is smaller for small $\sigma$.

Fig.~\ref{Fig::Mexhat_profile}(b) shows an oscillatory front profile for $\sigma=3$ within the parameter region between XR and LO. At difference from the BD transition here the oscillations appear at a finite spatial wavenumber and close to the HSS the front profile is very well described by a combination of two exponentials of the form
\begin{equation}
 A(x)-A_{st} \approx c_1 e^{q_1 x}+ c_2 e^{q_2 x} \cos(k_2 x + \phi_2) 
  \label{Eq::Front_fit_crossover}
\end{equation}
where $q_1=-0.518$ is the real spatial eigenvalue while $q_2=-0.6495$ and $k_2=0.734$ are the real and imaginary part of the complex spatial eigenvalue. The coefficients $c_1=-0.7014$, $c_2=0.9892$ and $\phi_2=-1.584$ have been fitted.

\section{Conclusions} \label{Sect::conclu}

In this manuscript we have applied the general framework developed in Part I (see Ref. \cite{PartI}) to 
illustrate the effect of nonlocal interactions using the GLE as a prototypical example. In particular 
the work presented here allows for a detailed explanation of some of the findings advanced in \cite{GelensPRL2010}.
One of the main results of \cite{GelensPRL2010} was that in spatially 
extended nonlinear systems exhibiting fronts connecting two equivalent 
homogeneous steady states, the addition of a spatially nonlocal linear 
interaction term can induce the creation of localized structures in systems with monotonic fronts. This interesting effect
is induced by a repulsive nonlocal interaction, that is able to
induce spatial tails, and, thus, lead to stable LSs. This was shown in Ref.
\cite{GelensPRL2010} for the case of a Gaussian nonlocal influence kernel.
Leveraging the general framework developed in Part I \cite{PartI}
we rationalize these results and to show its generality by considering
two other choices of the nonlocal influence kernel, a mod-exponential
and a Mexican-hat. Remarkably, in the case of the
two first kernels we have been able to find analytical conditions for
the existence of the LSs. In the case of the Mexican-hat kernel, with
coexisting attractive and repulsive interactions, LSs are obtained
through two different mechanisms for both the cases with short-range
excitation and long-range inhibition and the other way around. One mechanism is the Belyakov-Devaney  transition \cite{Devaney76,Homburg10} in which oscillations appear initially at infinite wavelength, and the other a crossover in the location of the spatial eigenvalues on the complex plain in which finite wavelength oscillations develop. 

There are a number
of problems that exhibit bistable dynamics and domain walls
connecting them and also, presumably, spatially nonlocal effects,
e.g., in chemical reactions \cite{Boissonade2006} and in nonlinear optics
\cite{Esteban_06,*Taranenko_98}. The present work shows that spatial nonlocal
effects can have a big influence on these phenomena. Comparisons
with experimental results can be made more quantitative by
reconstructing the experimental kernel \cite{Minovich,Hellwig}.

\section*{Acknowledgments}
This work was supported by the Belgian Science Policy Office (BeISPO) under grant No.\
IAP 7-35, and by the Spanish MINECO and FEDER 
under Grants FISICOS (FIS2007-60327), DeCoDicA (TEC2009-14101),
INTENSE@COSYP (FIS2012-30634), and TRIPHOP (TEC2012-36335), and
from Comunitat Aut\`onoma de les Illes Balears. LG acknowledges support by the
Research Foundation - Flanders (FWO). We thank Prof. E. Knobloch and Dr. G. Van der Sande for interesting discussions.

\appendix
\section{Lambert's W Function} \label{sec:Lambertap}

The so-called Lambert's W function \footnote{This special function is
quite accessible nowadays as it can be found in modern scientific programs
like Mathematica (where it is called {\it ProductLog}), Matlab (where it is called
{\it lambertw}, and Maple (where it is called
{\it LambertW}).} is the inverse function of $x=f(y)=y\exp(y)$, i.e., $y=W(x)$ (see, e.g.,
\cite{Corless96,MapleW} for further details). It
can be seen as a generalized algorithm, a useful analogy, because as the (complex)
logarithm function, Lambert's W function is multivalued. So, we will define it as,
\begin{equation}
x=W_l(x) \exp(W_l(x))\ ,\quad l\in \mathbb{Z}
\label{eq:lambertdef}
\end{equation}
where, in principle, $x\in \mathbb{C}$ and $l\in \mathbb{Z}$ is the branch index. The principal branch, $W_0(x)$
or simply $W(x)$, has a branch point at $x=-1/e$ and a branch cut along the negative
real axis $x\in [-\infty,-1/e]$ ($W_0(-1/e)=-1$), and is real valued in the interval $x\in [-1/e,\infty]$.
Moreover, it is analytic at $x=0$, $W_0(0)=0$, while all the other branches have a 
branch point at $0$. Moreover, $W_{-1}(x)$ is real in the interval $x\in [-1/e,0]$. 
   	
The Lambert function can also be used to find the exact solution of transcendental
equations of the type $x+\exp(x)=a$. 
Thus, a solution to the equation,
\begin{equation}
cx+\exp(ax)=b 
\end{equation}
can be found with the change,
\begin{equation}
\frac{y}{a}=b-c x
\end{equation}
as
\begin{equation}
\frac{y}{c}=W\left[\frac{a}{c}\exp(ab/c)\right]\\
\end{equation}
and undoing the change of variables one gets,
\begin{equation}
x=\frac{b}{c}-\frac{1}{a} W\left[\frac{a}{c}\exp(ab/c)\right]
\end{equation}

The real branches of Lambert's W function admit the following series expansions, valid
close to $x=0$ ($x<0$ for $W_{-1}(x))$,
\begin{eqnarray}
W_0(x)=\sum_{n=1}^{\infty} x^n=x-x^2+\frac{3}{2} x^3+\ldots \label{eq:powexpw0}\\
W_{-1}(x)=\ln(-x)-\ln(-\ln(-x))+\ldots
\label{eq:powexpwm1}
\end{eqnarray}

\begin{figure}[t]
\begin{center}
\includegraphics[width=8cm]{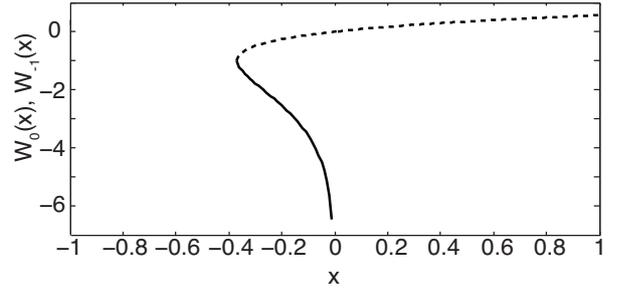}
\caption{\label{Fig::Lambertfunct} A plot of the real branches of Lambert's W function,
two-valued in the range $[-1/e,0]$.}
\end{center}
\end{figure}

\bibliography{Nonlocal}

\end{document}